\documentclass[prl,twocolumn,superscriptaddress]{revtex4-1}
\usepackage{epsfig}
\usepackage{graphicx}
\usepackage{palatino}
\usepackage[english]{babel}
\usepackage{hyphenat}
\usepackage{amsmath}
\usepackage{amssymb}
\usepackage{mathtools}
\usepackage{mathrsfs}
\usepackage{bbm} 
\usepackage{bm}
\usepackage{slashed}
\usepackage{epstopdf}
\usepackage{xcolor}
\usepackage{booktabs}
\usepackage{float}
\definecolor{lcolor}{rgb}{0.,0.0,0.}
\definecolor{citcolor}{rgb}{0,0.,0.5}
\usepackage[breaklinks,colorlinks,urlcolor=blue,citecolor=blue,linkcolor=blue]{hyperref}
\usepackage{multirow}
\usepackage{ltablex}
\usepackage{soul}

\newcommand{\rmT}{\textrm{T}}
\newcommand{\Hcal}{\mathcal{H}}

\newcommand{\vect}[1]{\boldsymbol{#1}_{\perp}}
\newcommand{\kt}{\vect{k}}
\newcommand{\kgt}{\boldsymbol{k_{g\perp}}}
\newcommand{\ktp}{k_{g\perp}}

\newcommand{\Pt}{\vect{P}}

\newcommand{\qt}{\vect{q}}

\newcommand{\bt}{\vect{b}}

\newcommand{\ktone}{\boldsymbol{k_{1\perp}}}

\newcommand{\ltthre}{\boldsymbol{l_{3\perp}}}

\newcommand{\xt}{\vect{x}}
\newcommand{\yt}{\vect{y}}
\newcommand{\zt}{\vect{z}}

\newcommand{\rt}{\vect{r}}

\newcommand{\rxyt}{\boldsymbol{r}_{xy}}

\newcommand{\rzxt}{\boldsymbol{r}_{zx}}
\newcommand{\rzyt}{\boldsymbol{r}_{zy}}

\newcommand{\rxxtp}{\boldsymbol{r}_{xx'}}

\newcommand{\ellt}{\boldsymbol{\ell}_\perp}

\newcommand{\rzxpt}{\boldsymbol{r}_{zx'}}

\newcommand{\rxpyt}{\boldsymbol{r}_{x'y}}

\newcommand{\RtS}{\boldsymbol{R}_{\rm SE}}
\newcommand{\RtV}{\boldsymbol{R}_{\rm V}}
\newcommand{\RtR}{\boldsymbol{R}_{\rm R}}

\newcommand{\der}{\mathrm{d}}

\newcommand{\Tr}{\mathrm{Tr}}

\newcommand{\eqn}[1]{Eq.\,\eqref{#1}}
\newcommand{\beq}{\begin{equation}}
\newcommand{\eeq}{\end{equation}}
\newcommand{\bal}{\begin{align}}
\newcommand{\eal}{\end{align}}

\newcommand{\del}{\partial}

\newcommand{\rmd}{{\rm d}}

\newcommand{\rme}{{\rm e}}
\newcommand{\bk}{\bm{k}}

\newcommand{\mcal}{\mathcal}

\newcommand{\PT}{P_\perp}

\newcommand{\RtRb}{\boldsymbol{R}_{\overline{\rm R}}}

\begin{document}

\title{Jet Definition and Transverse-Momentum-Dependent Factorization in Semi-Inclusive Deep-Inelastic Scattering}
\author{Paul Caucal}
\email{caucal@subatech.in2p3.fr}
 \affiliation{SUBATECH UMR 6457 (IMT Atlantique, Université de Nantes, IN2P3/CNRS), 4 rue Alfred Kastler, 44307 Nantes, France}
\author{Edmond Iancu}
\email{edmond.iancu@ipht.fr}
 \affiliation{Université Paris-Saclay, CNRS, CEA, Institut de physique théorique, F-91191, Gif-sur-Yvette, France}
\author{A. H. Mueller}
\email{ahm4@columbia.edu}
\affiliation{Department of Physics, Columbia University, New York, NY 10027, USA}
\author{Feng Yuan}
\email{fyuan@lbl.gov}
\affiliation{Nuclear Science Division, Lawrence Berkeley National Laboratory, Berkeley, CA 94720, USA}
\affiliation{Institute for Theoretical Physics,
                Universit\"{a}t T\"{u}bingen,
                Auf der Morgenstelle 14,
                D-72076 T\"{u}bingen, Germany}

\begin{abstract}
Using the colour dipole picture of Deep Inelastic Scattering (DIS) and the Colour Glass Condensate effective theory, we study  semi-inclusive jet production in DIS at small $x$ in the limit where the photon virtuality $Q^2$ is much larger than the transverse momentum squared $P_\perp^2$ of the produced jet. In this limit, the cross-section is dominated by aligned jet configurations, that is, quark-antiquark pairs in which one of the fermions --- the would-be struck quark in the Breit frame --- carries most of the longitudinal momentum of  the virtual photon. We show that physically meaningful jet definitions in DIS are such that the effective axis of the jet sourced by the struck quark is controlled by its virtuality rather than by its transverse momentum. For such jet definitions, we show that the next-to-leading order (NLO) cross-section admits factorisation in terms of the (sea) quark transverse momentum dependent (TMD) distribution, which in turn satisfies a universal Dokshitzer-Gribov-Lipatov-Altarelli-Parisi and Sudakov evolution.
\end{abstract} 

\maketitle

\paragraph{Introduction.} The semi-inclusive production of a hadron or a jet in Deep Inelastic Scattering (SIDIS) is a process of fundamental importance for the study of the partonic content inside a proton or a large nucleus. It allows for an unambiguous extraction of the quark and gluon distribution functions with increasing accuracy in perturbative QCD~\cite{Altarelli:1979kv,Baier:1979sp,deFlorian:1997zj,Abelof:2016pby,Goyal:2023zdi,Bonino:2024qbh}. In the small transverse momentum region of the measured hadron or jet, it is sensitive to the transverse momentum dependent (TMD) quark distribution
~\cite{Collins:1992kk,Ji:2004wu,Collins:2011zzd,Boussarie:2023izj}, whose precise extraction for both unpolarised and polarised target is one of the main goal of the Electron-Ion Collider (EIC) physics program~\cite{Accardi:2012qut,AbdulKhalek:2021gbh,Achenbach:2023pba}. In this Letter, we consider the SIDIS process in the case of a jet measurement at small $x_{\rm Bj}$, where the Colour Glass Condensate (CGC) effective theory \cite{Iancu:2002xk,Iancu:2003xm,Gelis:2010nm,Kovchegov:2012mbw} applies, and demonstrate that the factorisation of the NLO cross-section in terms of the (sea) quark TMD necessitates the use a new jet reconstruction algorithm which captures the non-trivial dynamics of the quark struck by the virtual photon. 

Compared to hadron production in SIDIS, jet production has the advantage that it can directly probe the TMD quark distribution without involving a TMD fragmentation function. However, a precise measurement of low transverse momentum jets is very challenging, in particular because the usual $k_t$-type algorithms may not be applicable. Alternative approaches and detailed phenomenology applications have been carried out in the last few years~\cite{Gutierrez-Reyes:2018qez,Gutierrez-Reyes:2019msa,Gutierrez-Reyes:2019vbx,Liu:2018trl, Liu:2020dct,Aschenauer:2019uex,Page:2019gbf,Arratia:2019vju,Arratia:2020ssx,Arratia:2020nxw,Kang:2021ryr,Kang:2020fka,Kang:2021ffh,H1:2021wkz,Lai:2022aly,Lee:2022kdn,Arratia:2022oxd,Fang:2023thw}. Together with these developments, the novel jet algorithm proposed in this paper will help to improve the precision of future jet measurements in SIDIS at the EIC and provide important probe to the nucleon/nucleus tomography in terms of the TMD quark distributions at various $x$ ranges. 

\begin{figure}[t]
    \centering
    \includegraphics[width=0.48\textwidth]{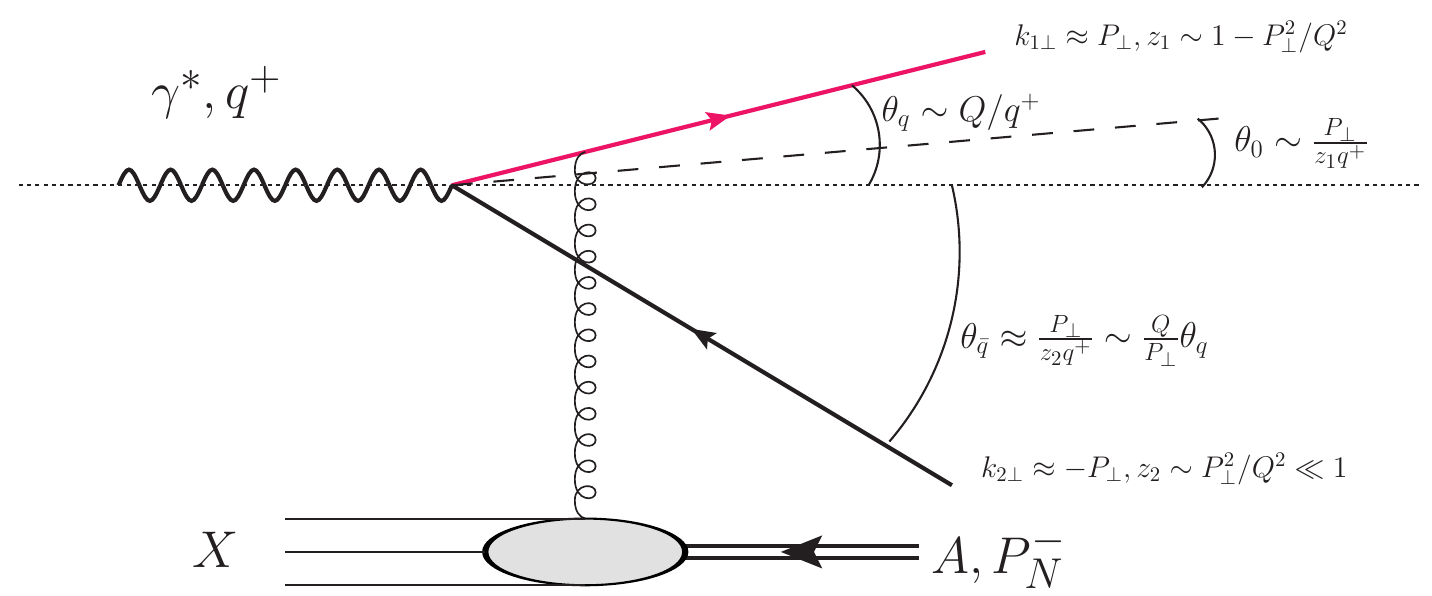}
    \caption{Geometric representation of the typical aligned jet configuration at small $x_{\rm Bj}$ in the dipole frame.}
    \label{fig:qqbar-angles}
\end{figure}

\paragraph{Aligned jet configurations.} We work in the dipole frame (related to the Breit frame by a longitudinal boost) where the virtual photon has four-momentum $q^\mu=(q^+,q^-=-Q^2/(2q^+),\boldsymbol{0}_\perp)$ and a nucleon from the nucleus target has four momentum $P^\mu_N=(0,P^-_N,\boldsymbol{0}_\perp)$ in light-cone coordinates. The standard DIS variables are defined by $Q^2=-q^2$ and $x_{\rm Bj}=Q^2/\hat s$ with $\hat s=2q^+P^-_N$.  At LO in the colour dipole picture of DIS at small $x_{\rm Bj}$~\cite{Kopeliovich:1981pz, Bertsch:1981py, Mueller:1989st,Nikolaev:1990ja}, the transversely polarised virtual photon $\gamma_{\rmT}^*$ splits into a quark-antiquark pair which interacts with the nucleus target. We denote by $k_{1,2}^\mu$ the four-momenta of the quark and the antiquark and define $z_i=k_i^+/q^+$. Each outgoing parton can then form a jet for which one only measures the transverse momentum $\Pt$. We are primarily interested in the limit $Q^2\gg P_\perp^2$. 
The leading power in the $P_\perp^2/Q^2$ expansion of the cross-section comes from aligned-jet configurations~\cite{Mueller:1999wm,Marquet:2009ca,Iancu:2020jch}, that is very asymmetric $q\bar q$ pairs such that the quark has $1-z_1\sim P_\perp^2/Q^2\ll 1$ and carries most of the longitudinal momentum of the virtual photon while the antiquark has $z_2\sim P_\perp^2/Q^2 \ll 1$, or vice-versa (cf.\,Fig.\,\ref{fig:qqbar-angles}). The struck fermion in the target picture is naturally the one with $z_i\sim 1$, yet the tagged jet with transverse momentum $\Pt$ can be either the fast ($z_i\sim 1$),  or the slow  ($z_i\ll 1$),  fermion in the dipole picture. In the Breit frame, this would give either a forward jet in the direction of the photon, or a backward jet close to the beam remnants.

In the limit $Q^2\gg P_\perp^2$, the semi-inclusive single jet cross-section admits TMD factorisation~\cite{Marquet:2009ca} in terms of the sea quark TMD $x \mathcal{F}_q(x, \Pt)$ at small $x$~\cite{McLerran:1998nk,Venugopalan:1999wu,Mueller:1999wm}:
\begin{align}
    \left.
    \frac{\der\sigma^{\gamma^*_{\rm T}+A\to j+X}}{\der^2\Pt}\right|_{\rm LO}&=\frac{8\pi^2\alpha_{\rm em}e_f^2}{Q^2} x\mathcal{F}_{q}^{(0)}(x, \Pt)\,,\label{eq:LO-SIDIS}
\end{align}
where $x=x_{\rm Bj}$ as a consequence of minus momentum conservation. The superscript $(0)$ for $\mathcal{F}_q$ refers to a LO approximation for the quark TMD including its high energy BK/JIMWLK evolution~\cite{Balitsky:1995ub,Kovchegov:1999yj,JalilianMarian:1997jx,JalilianMarian:1997gr,Kovner:2000pt,Weigert:2000gi,Iancu:2000hn,Iancu:2001ad,Ferreiro:2001qy} down to the value of $x$ of interest. $\alpha_{\rm em}$ is the fine structure constant and $e_f$ is the fractional electric charge of a light quark. This expression covers the case where the quark jet is measured~\footnote{A similar expression holds for the case where the antiquark jet is measured. See \cite{Cacciari:2019qjx} for a discussion of the unitarity of the SIDIS cross-section with jets.}, but it is inclusive in the jet longitudinal momentum fraction $z$. 
Explicit expression for the sea quark TMD in term of the quark-antiquark dipole $S$-matrix at small $x$ can be found in~\cite{Xiao:2017yya,Marquet:2009ca,Tong:2022zwp}
(see also the Supplemental Material~\footnote{See Supplemental Material, which includes Refs.~\cite{Newman:2013ada,ZEUS:2002nms,ZEUS:2006xvn, ZEUS:2010vyw,H1:2007xjj,H1:2016goa,ZEUS:2023zie,Liu:2013hba,Catani:1993hr,Blazey:2000qt,Ducloue:2019jmy,Dominguez:2011wm,Bergabo:2022tcu,Iancu:2022gpw,Bergabo:2023wed,Bergabo:2022zhe,Becher:2010tm,Becher:2012yn,Echevarria:2011epo,Echevarria:2014rua,Chiu:2012ir,Ebert:2019tvc,Sun:2013hua}, for a comparison of the various jet definitions used in DIS, details on the derivation of Eq.\,\eqref{eq:virtual-xs} and a general discussion on TMD factorisation for jet production in SIDIS at moderate values for $x$.} for a brief review).

\paragraph{Heuristic discussion of the Sudakov logarithm.} At NLO in the strong coupling $\alpha_s$, the dominant radiative corrections to Eq.\,\eqref{eq:LO-SIDIS} are enhanced by double Sudakov logarithms~\cite{Sudakov:1954sw,Dokshitzer:1978hw,Parisi:1979se,Collins:1984kg} in the ratio of $Q^2/P_\perp^2$. To DLA, these Sudakov logarithms come from virtual emissions by the struck quark in the phase-space that is forbidden to the real gluon emissions, that is real gluons which would modify the structure of the final state~\cite{Mueller:2012uf,Mueller:2013wwa}. Clearly, the real gluon emissions with transverse momenta such that $\ktp^2\gg P_\perp^2$ must be forbidden. On the other hand, virtual emissions with $\ktp^2\ge Q^2$ do not have any double logarithmic support (the UV logarithmic divergence cancels above the hard scale $Q^2$ among virtual graphs~\cite{Caucal:2024cdq}). The transverse phase-space which contributes to the Sudakov double logarithm is therefore $P_\perp^2\ll \ktp^2\ll Q^2 $. Concerning the longitudinal phase space, we observe that the gluon emission must have a formation time $\tau_g=1/k_g^-$ larger than the coherence time $\tau_\gamma=1/|q^-|$ of the virtual photon \cite{Beuf:2014uia,Iancu:2015vea,Hatta:2016ujq,Ducloue:2019ezk}. In the context of high energy factorisation, this condition is imposed on any gluon emission which does not contribute to the (collinearly improved) Balitsky-Kovchegov (BK) and Jalilian-Marian-Iancu-McLerran-Weigert-Leonidov-Kovner (JIMWLK) evolutions~\cite{Balitsky:1995ub,Kovchegov:1999yj,JalilianMarian:1997jx,JalilianMarian:1997gr,Kovner:2000pt,Weigert:2000gi,Iancu:2000hn,Iancu:2001ad,Ferreiro:2001qy} of the target wave-function. In terms of $z_g=k_g^+/q^+$, this gives the constraint $z_g\gg \ktp^2/Q^2$. We thus end up with the following integral for the Sudakov double logarithm,
\begin{align}
S_{\rm DL}=-\frac{\alpha_sC_F}{2\pi}\int_{ P_\perp^2}^{Q^2}\frac{\der \ktp^2}{\ktp^2}\int_{\ktp^2/Q^2}^{z_{\rm max}}\frac{\der z_g}{z_g}\,,\label{eq:DL-phasespace}
\end{align}
which factorises from the LO cross-section Eq.\,\eqref{eq:LO-SIDIS}.
 
Consider now the upper limit $z_{\rm max}$ of the $z_g$ integral. In the case of a {\it hadron} measurement, a forbidden real gluon must be well separated from the collinear singularity for a final-state emission. This singularity corresponds to (soft) gluons obeying $\kgt=z_g\Pt/z_1$, with $z_1\simeq 1$ for the struck quark. Clearly such collinear gluons have no overlap with the forbidden phase-space at $\ktp^2\gg P_\perp^2$, so the collinear singularity introduces no additional constraint on this phase-space. We can then take $z_{\rm max}=1$, which gives 
\begin{align}
S_{\rm DL}^{\rm had}=-\frac{\alpha_sC_F}{2\pi}\ln^2\left(\frac{Q^2}{P_\perp^2}\right),
\label{eq:Sudakov-DL-had}
\end{align}
 as expected for hadron production~\cite{Xiao:2017yya,Altinoluk:2024vgg}. 

The case of a {\it jet} measurement is considerably more subtle, and here comes our main physical observation. The fast quark with $z_1\sim1$ (see Fig.\,\ref{fig:qqbar-angles}) is not put on-shell by the scattering, rather it emerges from 
the collision with a relatively large virtuality, of order $Q^2$.
 Indeed, after the photon decay $\gamma\to q\bar q$, the two quarks separate via quantum diffusion. 
 By the time $\tau_\gamma =2q^+/Q^2$ of
the collision with the target, the wave packets of the fast quark and the slow antiquark spread out over distances $\Delta x_{q\perp}^2\sim \tau_\gamma/(z_1q^+)$ and $\Delta x_{\bar q\perp}^2\sim \tau_\gamma/(z_2q^+)$, respectively. In the aligned jet limit, $\Delta x_{q\perp}^2\sim 1/Q^2$ and $\Delta x_{\bar q\perp}^2\sim1/P_\perp^2$. For the fast quark,   $\Delta x_{q\perp}^2\ll 1/P_\perp^2$ is very small, showing that this parton is still localised by its virtuality. Physically, this is so since 
the collision occurs relatively fast:  
$ \tau_\gamma=2q^+/Q^2$  is much smaller than the formation time $\tau_q\sim 2z_1q^+/P_\perp^2$
for the fast quark. Accordingly, the angle made by  this quark with respect to the collision 
axis by the time of scattering can be
 estimated as $\theta_q\sim \Delta x_{q\perp}/\tau_\gamma\sim Q/q^+$, which is much bigger than the naive angle 
 $\theta_0\sim P_\perp/q^+$ it would have made if it was on-shell (i.e.\, localised by its transverse momentum, see Fig.\,\ref{fig:qqbar-angles}).

In order to be forbidden, a real gluon emission must not be part of the jet sourced by the fast quark. This condition imposes $\theta_g\sim k_{g\perp}/k_g^+\gg \theta_q$, i.e.\,$z_g\ll k_{g\perp}/Q$.  
Using $z_{\rm max}=k_{g\perp}/Q$ in Eq.\,\eqref{eq:DL-phasespace}, one finds
\begin{align}
    S_{\rm DL}^{\rm jet}&=-\frac{\alpha_sC_F}{4\pi}\ln^2\left(\frac{Q^2}{P_\perp^2}\right)\,.\label{eq:Sudakov-DL-jet}
\end{align}
Thus, remarkably, the Sudakov corresponding to a jet final state is smaller by a factor of 2 than that for a hadron
final state. This is so since the fast virtual quark generates a relatively wide jet, so the phase-space that is forbidden to real emission is correspondingly smaller. 

\paragraph{A new jet distance measure in DIS.} To properly reflect the above physical picture, a jet measurement in DIS must be endowed with a clustering algorithm which accounts for the high virtuality $\sim Q^2$ of the struck quark. This is generally not the case for the jet algorithms that are a priori designed for hadron-hadron collisions, such as the $k_t$-algorithms \cite{Ellis:1993tq,Dokshitzer:1997in,Wobisch:1998wt,Cacciari:2008gp,Cacciari:2011ma}: when applied to DIS, they typically cluster particles "around" an axis with angle $P_\perp/(zq^+)$ in the dipole frame~\cite{Ivanov:2012ms,Boussarie:2016ogo,Roy:2019cux,Roy:2019hwr,Caucal:2021ent,Taels:2022tza,Caucal:2022ulg,Liu:2022ijp} and thus do not capture the effect of the large virtuality of the jet with $z\sim 1$.  Related to that, they fail to cluster the remnant of the struck quark into the same jet, as noted in \cite{Arratia:2020ssx}. In order to cope with these issues, we introduce a new jet distance measure, via
\begin{align}
    d_{ij}=\frac{M_{ij}^2}{(z_iz_j)^pQ^2R^2}\,,\quad d_{iB}=1\,,\label{eq:jet-def}
\end{align}
where $d_{ij}$ is the distance between two particles labelled $i,j$, and $d_{iB}$ is the particle-beam distance~\footnote{The measure corresponds to the particular case $k=0$ of a generalised distance measure $d_{ij}=\textrm{min}(k_{i\perp}^{2k},k_{j\perp}^{2k})M_{ij}^2/((z_iz_j)^pQ^2R^2)$, $d_{iB}=k_{i\perp}^{2k}$ as done in hadron-hadron collisions~\cite{Cacciari:2008gp,Salam:2010nqg}.}. The measure depends on the invariant mass squared $M_{ij}^2=(k_i+k_j)^2$ and the longitudinal momentum fractions~$z_i=k_i\cdot P_N/(q\cdot P_N)$. For a given list of final state particles, the jet algorithm then runs inclusively~\cite{Salam:2010nqg} through pairwise recombination of particles $i_0,j_0$ 
if $d_{i_0j_0}$ is minimal among all $d_{ij}$,$d_{iB}$, $d_{jB}$, while declaring $i_0$ a final state jet if $d_{i_0B}$ is minimal, until no particle remains.

The algorithm depends on two parameters, the jet radius $R$ and the power $p$. With $p=-1,0,1$, it gives respectively a $k_t$, mass, or angular ordered clustering in the dipole frame.
We shall focus here on the specific choice $p=1$, which is dynamically favoured, as we shall shortly argue. With this choice, one finds (in the dipole frame
and in the limit of a small relative angle $\theta_{ij}\ll 1$)
\begin{align}
    d_{ij}\approx \frac{\theta_{ij}^2 (q^+)^2}{Q^2R^2}\,,\quad \textrm{ for }p=1\,.\label{eq:dij-smalltheta}
\end{align}
Using this criterion at NLO, where the final state is made of two partons, the quark $i$ and gluon $j$, one sees that these two partons are
clustered within the same jet if $\theta_{ij}\le R \theta_q$ with $\theta_q\sim Q/q^+$, in agreement with the dynamics of the fast virtual quark that we have just elucidated. This is a consequence of both the normalisation of the distance measure with $Q^2$ and the choice of $p$. We assume the jet radius $R$ to be small, but not {\it too} small,
so that an additional $\ln(1/R)$ resummation is not needed~\cite{Dasgupta:2014yra}~\footnote{For jet production with a very small-$R$, one needs to introduce the so-called semi-inclusive jet function~\cite{Kang:2016mcy} and the factorisation will be different from what we discuss in this paper.}.

For $p=1$ and in the fragmentation region of the struck quark in the Breit frame, the algorithm is akin to generalised $k_t$ algorithms originally designed for $e^+e^-$ annihilation~\cite{JADE:1986kta,JADE:1988xlj,Catani:1991hj,Cacciari:2011ma} and subsequently extended to DIS in~\cite{Catani:1992zp,Webber:1993bm} 
(cf Supplemental Material). However,  unlike the latter, which are only defined
in the Breit frame, Eq.\,\eqref{eq:jet-def} is longitudinally invariant by boost along the $\gamma^*$-$A$ collision axis (see also~\cite{Arratia:2020ssx,vanBeekveld:2023chs}), so that $d_{ij},d_{iB}$ can be computed either in the Breit frame, or the dipole frame.
This makes the distance measure Eq.\,\eqref{eq:jet-def} more convenient for higher order calculations and jet studies in DIS.

\paragraph{TMD factorisation at NLO.} For the \textit{inclusive} production of jets in DIS, that is, in situations where the jet transverse momenta are not measured, but only their number, it was known since~\cite{Catani:1992zp,Webber:1993bm} that special jet definitions are needed to guarantee the validity of the standard collinear factorisation for the jet  cross-section. In what follows, we would like to demonstrate a similar property for the semi-inclusive production of a jet with a given $P_\perp$. Specifically, we will show that the NLO corrections to the jet cross-section (as computed in the  dipole picture) are consistent with TMD factorisation --- in the sense that they correctly
generate the expected Dokshitzer-Gribov-Lipatov-Altarelli-Parisi (DGLAP)~\cite{Gribov:1972ri,Altarelli:1977zs,Dokshitzer:1977sg}  and Collins-Soper-Sterman (CSS)~\cite{Collins:1981uk,Collins:1981uw,Collins:1984kg,Collins:2011zzd} evolutions of the quark TMD --- if and only if the quark and the gluon jets are
separated from each other by using the jet distance \eqref{eq:jet-def}  with $p=1$. 
To show that, we consider a relatively hard jet, with  $Q^2  \gg P_\perp^2\gg Q_s^2$. In this regime and to lowest order
in $ Q_s^2/P_\perp^2$, the transverse momentum of the measured jet is either given by the recoil from a hard gluon emission with $\kgt\sim -\Pt$ (real contribution) or by the target itself (virtual contribution). 

For the real contribution, our starting point is the quark-antiquark-gluon Fock component of the virtual photon wavefunction and,
more precisely, its contribution to the $\gamma_{\rm T}^*+A\to qg$ cross-section, as obtained after integrating out
the antiquark~\footnote{If instead of the antiquark, one integrates out the gluon, that is, if one computes a real-gluon correction to the $\gamma_{\rm T}^*+A\to qq$ cross-section, then one generates a step in the 
DGLAP+CSS evolution of the {\it gluon} TMD~\cite{Caucal:2024bae}.}.  The quark and the gluon are nearly back-to-back:
their transverse momentum imbalance $\ellt=\kgt+\ktone$ is small compared to the  relative momentum
$\Pt=z_g\ktone-z_1\kgt$:  $\ell_\perp\ll P_\perp$ and $z_1+z_g\simeq 1$. In the CGC calculation, this process is shown to factorise in terms of the quark TMD as~\cite{Hauksson:2024bvv,Caucal:2024xxx}
\begin{align}
   &\frac{\der\sigma^{\gamma_{\rmT}^\star+A\to q g+X}}{\der^2\Pt\der^2\ellt\der z_1\der z_g}=\alpha_{\rm em}e_f^2\alpha_s C_F\delta(1-z_1-z_g) \nonumber\\
   &\times\frac{2z_1\left[( P_\perp^2+\bar Q^2)^2+z_g^2P_\perp^4+z_1^2\bar Q^4\right]}{ P_\perp^2\left[ P_\perp^2+\bar Q^2\right]^3}x_q \mathcal{F}_q^{(0)}(x_q,\ellt)\,,\label{eq:qg-xs}
\end{align}
with $\bar Q^2=z_1z_g Q^2$ and $x_q=(M_{qg}^2+Q^2)/\hat s$. The contribution of \eqref{eq:qg-xs} to SIDIS is obtained
by integrating out the kinematics of the unmeasured jet over the region in phase-space where the quark and the gluon form two 
separated jets. Using the jet definition Eq.\,\eqref{eq:jet-def} with $p=1$, the associated phase space constraint reads $P_\perp^2\ge R^2Q^2z_1^2z_g^2$. The subsequent integrals over $z_1$ and $z_g$ are controlled by $z_1\simeq 1$ 
and  $ z_g\lesssim P_\perp/(R Q)\ll 1$,  so only relatively soft gluons contribute.
Such soft gluons can be transferred into the target wavefunction~\cite{Iancu:2021rup,Iancu:2022lcw,Hauksson:2024bvv,Caucal:2024bae} via a change of variable $z_g\to\xi $,  with 
\begin{equation}
    z_g=\frac{\xi}{1-\xi}\frac{P_\perp^2}{Q^2}\ \ \Longleftrightarrow \ \ \xi=\frac{x_{\rm Bj}}{x_q}
\end{equation}
where we have also used $M_{qg}^2=P_\perp^2/(z_1z_g)$ and $z_1\simeq 1$. 
The integral over $\xi$ is restricted to
$x_{\rm Bj}< \xi < 1-RP_\perp/Q$, where the upper limit  is introduced by the  jet constraint $z_g\lesssim P_\perp/(R Q)$, while the lower limit comes from the condition $x_q\le1$. Note that Eq.~\eqref{eq:qg-xs} is not singular as $z_g\to 0$  meaning that the contribution from Eq.~\eqref{eq:qg-xs} to the NLO cross-section does not overlap with the high energy evolution already accounted for through the $x$ dependence of $x\mathcal{F}_q^{(0)}$ in Eq.~\eqref{eq:LO-SIDIS}.
Finally, the integral over the momentum imbalance
$\ell_\perp\ll P_\perp$ builds the quark PDF $xf_q$ on the resolution scale of the hard jet:
\begin{align}
    x f_q^{(0)}(x,P_\perp^2)=\int_{\Lambda^2}^{P_\perp^2} {\der^2 \ellt} \,x\mathcal{F}_q^{(0)}(x,\ellt)\,,\label{eq:PDF}
\end{align}
with $\Lambda$ of the order of the QCD confinement scale. Based on the above, a straightforward calculation yields
\begin{align}
     &\left.\frac{\der\sigma^{\gamma^*_{\rm T}+A\to j+X}}{\der^2\Pt}\right|_{R}= \frac{8\pi^2\alpha_{\rm em}e_f^2}{Q^2}\nonumber\\
     &\times \frac{\alpha_s}{2\pi^2}\frac{1}{P_\perp^2}\int_{x_{\rm Bj}}^{1-\frac{RP_\perp}{Q}} {\der \xi} P_{qq}(\xi)\frac{x_{\rm Bj}}{\xi}f_q^{(0)}\left(\frac{x_{\rm Bj}}{\xi},P_\perp^2\right)\,,\label{eq:real-DGLAP}
\end{align}
with $P_{qq}(\xi)=C_F(1+\xi^2)/(1-\xi)$ the unregularised $q\to qg$ splitting function. In
\eqn{eq:real-DGLAP} we included a factor of two to account for the fact that the tagged jet can be generated by 
either the fast quark ($z_1\simeq 1$), or the slow gluon ($z_g\ll 1$).  The NLO correction in \eqn{eq:real-DGLAP} exhibits
TMD factorisation, as anticipated: the expression in the second line can be interpreted as an evolution of the quark TMD.
The singularity of $P_{qq}(\xi)$ at $\xi=1$
introduces a logarithmic sensitivity to the upper limit, that can be isolated with the help of the plus prescription:
\begin{align}
\hspace{-0.8cm}  
& x\mathcal{F}_{q}^{(1)}(x,\Pt, Q^2)\Big|_{R}=
     \frac{\alpha_s}{2\pi^2}\frac{1}{P_\perp^2}\!\int_{x}^{1}\! {\der \xi} 
     P_{qq}^{(+)}(\xi)\frac{x}{\xi}f_q^{(0)}\!\left(\frac{x}{\xi},P_\perp^2\right)
     \nonumber\\
    &  
      \quad \qquad\qquad  + \frac{\alpha_s C_F}{2\pi^2}\frac{1}{P_\perp^2}\ln \left(\frac{Q^2}{R^2 P_\perp^2}\right)
        x f_q^{(0)}(x,P_\perp^2)
    \,,\label{eq:real-smallx}
\end{align}
where $P_{qq}^{(+)}(\xi)$ differs from $P_{qq}(\xi)$ only via  the replacement  $(1-\xi)\to (1-\xi)_+$ in the denominator.
We recognise in Eq.\,\eqref{eq:real-smallx} one (real) step in the DGLAP+CSS evolution of the quark TMD.
(See also \cite{Hauksson:2024bvv,Caucal:2024bae} for similar arguments.) As shown in the SM, Eq.\,\eqref{eq:real-smallx} is also the standard result for the one-loop contribution of the collinear gluon radiation to the quark TMD at moderate $x$ and $P_\perp\gg \Lambda$~\cite{Boussarie:2023izj}. Recovering this well known result is a non-trivial check of our jet definition. Indeed, the precise coefficient of the Sudakov logarithm in the second line is a consequence of choosing $p=1$ in our jet distance measure \eqref{eq:jet-def}.   If one were using another jet definition from that class,
 or a definition with an effective jet axis set by the angle $\theta_0$ in Fig.\,\ref{fig:qqbar-angles}, the upper limit of the $\xi$ integral in Eq.\,\eqref{eq:real-DGLAP} would change (e.g., it would be $\xi<1-R^2 P_\perp^2/Q^2$ for $p=0$), which would in turn modify the normalisation of the Sudakov logarithm. With
$p=-1$ in \eqn{eq:jet-def}, the quark and the gluon would typically be clustered within the same jet, 
so that the TMD evolution that we have just unveiled would not be resolved within the wide jet formed by the struck quark.

An important consistency check refers to the detailed balance between the real and the virtual NLO corrections:
after integrating the cross-section over $\Pt$,  the Sudakov logarithms must cancel between real and virtual terms
(see e.g. the discussion in \cite{Caucal:2024bae}). In order to verify this condition and also to complete our calculation
of the cross-section to the accuracy of interest, we need the virtual NLO contributions in the limit $Q^2\gg P_\perp^2\gg Q_s^2$. They can be inferred from the NLO calculation of the SIDIS cross-section in the CGC, as presented in~\cite{Caucal:2024cdq}. The details of the leading power extraction in these expressions are provided in Supplemental Material. The virtual terms too are sensitive to the jet definition, as clear from the fact that they include the phase-space region
where the  quark and the gluon are nearly collinear with each other and hence must be clustered within the same jet. 
The collinear singularity cancels between real and virtual contributions, but the finite reminder, which is enhanced
by (double and single) Sudakov logarithms, is clearly dependent upon our definition for the jets. Using \eqn{eq:jet-def} with $p=1$, the virtual term reads
\begin{widetext}
\begin{align}
  \left.\frac{\der\sigma^{\gamma^*_{\rm T}+A\to j+X}}{\der^2\Pt}\right|_{V}=\left.\frac{\der\sigma^{\gamma^*_{\rm T}+A\to j+X}}{\der^2\Pt}\right|_{\rm LO}& \times  \frac{\alpha_sC_F}{\pi}\left[-\frac{1}{4}\ln^2\left(\frac{Q^2}{P_\perp^2}\right)+\left(\frac{3}{4}+\ln(R)\right)\ln\left(\frac{Q^2}{P_\perp^2}\right)\right.\nonumber\\
  &\left.-\frac{3}{2}\ln(R)+\frac{11}{4}-\frac{3\pi^2}{4}+\frac{3}{4}\ln^2(x_\star)+\frac{3}{8}\ln(x_\star)+\mathcal{O}(R^2)\right]\,.\label{eq:virtual-xs}
\end{align}
\end{widetext}
where $x_\star$ is a $\mathcal{O}(1)$ number that marks the separation between the phase spaces $z_g\le x_\star k_{g\perp}^2/Q^2$ contributing to collinearly improved BK/JIMWLK evolution  \cite{Beuf:2014uia,Iancu:2015vea,Hatta:2016ujq,Ducloue:2019ezk} of the quark TMD, and $z_g\ge x_\star k_{g\perp}^2/Q^2 $ contributing to the NLO impact factor. Since the latter has undergone a power expansion in $P_\perp/Q$ and $Q_s/P_\perp$, unlike the former, the cancellation of the $x_\star$ dependence between small $x$ evolution and NLO impact factor is not complete anymore, but the remaining $x_\star$ dependence is a pure NLO effect, as clear from Eq.~\eqref{eq:virtual-xs}.

 The first term in the square bracket in Eq.\,\eqref{eq:virtual-xs} is the Sudakov double logarithm that we heuristically derived in Eq.\,\eqref{eq:Sudakov-DL-jet}. It agrees with the expectation from the CSS kernel for the quark TMD alone~\cite{Collins:1981uk,Collins:1981uw,Collins:1984kg,Collins:2011zzd}.  The second term is a Sudakov single logarithm, which depends both on the quark anomalous dimension $\Gamma_q=\frac{3\alpha_sC_F}{4\pi}$ and on the jet parameter $R$. By only keeping these logarithmic terms,
 one obtains the virtual contribution to the DGLAP+CSS evolution of the quark TMD:
 
 \begin{align}
 &  \left.x\mathcal{F}_{q}^{(1)}(x,\Pt, Q^2)\right|_{V}=- \frac{\alpha_s C_F}{2\pi}
  \,x\mathcal{F}_q^{(0)}(x,\Pt)
 \nonumber\\
& \quad\quad \times  \left[\frac{1}{2}\ln^2\left(\frac{Q^2}{P_\perp^2}\right)-\left(\frac{3}{2}+\ln(R^2)\right)\ln\left(\frac{Q^2}{P_\perp^2}\right)\right].
  \label{eq:virt-CSS}
\end{align}
After integrating Eqs.\,\eqref{eq:real-smallx} and \eqref{eq:virt-CSS} over $ P_\perp^2$ up to $Q^2$, it is straightforward to verify that the effects of  the Sudakov logarithms which are manifest in these two equations mutually cancel, as announced. Furthermore, the piece  $3/2$ in \eqn{eq:virt-CSS} gives a contribution $(3/2)\delta(1-\xi)$ to the splitting function in Eq.\,\eqref{eq:real-smallx}, thus completing the standard expression for the regularised DGLAP splitting function $\mcal{P}_{qq}(\xi)$ \cite{Collins:2011zzd}. This real vs.~virtual  cancellation is a crucial feature of the jet definition that we employed, i.e.~\eqn{eq:jet-def} with $p=1$ (we show in Supplemental Material that other common choices of DIS jet definitions do not satisfy this condition). We thus obtain the DGLAP equation for the quark PDF in integral form:
\begin{align}
  x f_q(x,Q^2)  = & x f_q^{(0)}(x,Q^2)+ \int_{\Lambda^2}^{Q^2} \!\frac{\der P_\perp^2}{P_\perp^2}\frac
  {\alpha_s(P_\perp^2)}{2\pi}
     \nonumber\\
& \times \!\int_{x}^{1} {\der \xi} \,
     \mcal{P}_{qq}(\xi)\,\frac{x}{\xi}f_q\left(\frac{x}{\xi},P_\perp^2\right)
             ,\label{eq:DGLAP}
\end{align}
where we have also inserted a running coupling, as standard in this context. The CSS equation obeyed by the quark TMD
can be  obtained by taking a derivative w.r.t. $\ln Q^2$ in the sum of Eqs.\,\eqref{eq:real-smallx} 
and \eqref{eq:virt-CSS}. The precise relation between our ``top-down'' approach~\cite{Caucal:2024bae} 
to the resummation of Sudakov logarithms and the standard CSS formalism is given by $x\mathcal{F}_q(x,\Pt,Q^2)=x\mathcal{F}_q^{\rm (sub)}(x,\Pt,\mu_F=Q,\zeta_c=Q)$ where $\mathcal{F}_q^{\rm (sub)}$ is the subtracted quark TMD in Collins-11 scheme \cite{Collins:2011zzd} and $\mu_F$, $\zeta_c$ are respectively the UV and rapidity renormalisation scales, both identified with the hard scale $Q$ of the process~\footnote{For $\zeta_c=Q$, our longitudinal upper cut-off in $\xi\le 1-P_\perp/Q$ introduced the jet constraint is equivalent to the rapidity cut-off $y_n$ in Collins-11 scheme defined as $\zeta_c^2=2(x_{\rm Bj}P_N^-)e^{2y_n}$.}. As shown in \cite{Ebert:2022cku,delRio:2024vvq}, this diagonal scheme also has the advantage of preserving the validity of Eq.~\eqref{eq:PDF} connecting the quark TMD to the quark PDF (with $x\mathcal{F}_q^{(0)}\to x\mathcal{F}_q(x,\ellt,P_\perp^2)$) up to $\mathcal{O}(\alpha_s^2)$ corrections.

In the end, the single inclusive jet cross-section at small $x$ and $Q^2\gg P_\perp^2\gg Q_s^2$, for our new jet distance measure, can be written
\begin{align}
    \left.
    \frac{\der\sigma^{\gamma^*_{\rm T}+A\to j+X}}{\der^2\Pt}\right|_{\rm NLO}&=\frac{8\pi^2\alpha_{\rm em}e_f^2}{Q^2} x\mathcal{F}_{q}(x, \Pt,Q^2)\nonumber\\
    &\hspace{-0.5cm}\times\left[1-\frac{3\alpha_sC_F}{2\pi}\ln(R)+\mathcal{O}(\alpha_s)\right]\,,\label{eq:NLO-SIDIS-final}
\end{align}
where all the potentially large logarithms $\alpha_s\ln(1/x)$, $\alpha_s\ln(P_\perp^2/\Lambda^2)$,
 $\alpha_s\ln^{2}(Q^2/P_\perp^2)$  and  $\alpha_s\ln(Q^2/P_\perp^2)$ are resummed within the quark TMD via BK/JIMWLK ---  or Balitsky-Fadin-Kuraev-Lipatov (BFKL)~\cite{Kuraev:1977fs,Balitsky:1978ic} in the dilute limit $P_\perp \gg Q_s$ --- and DGLAP+CSS evolution equations, respectively.

To summarise, we have demonstrated that TMD factorisation for jet production in SIDIS is not guaranteed by all jet definitions and we have designed a longitudinally-invariant jet clustering algorithm which preserves both the factorisation and the universality of the DGLAP+CSS evolution of the quark TMD at small $x$. Physically, this jet definition is able to resolve the DGLAP and Sudakov dynamics of the struck sea quark and to distinguish the backward antiquark jet (in the Breit frame) from the beam remnant. While these results are derived in the context of the high energy factorisation, they also apply at moderate values of $x$, as demonstrated in the Supplemental Material, where we find a similar factorisation property by following the dynamics of the struck quark in the target picture. Once again, the proper choice of the jet definition turns out to be essential to ensure the correct matching between the NLO corrections to SIDIS and the DGLAP+CSS evolution of the quark TMD. It will be important to investigate further the jet definition
Eq.\,\eqref{eq:jet-def} both at small and moderate $x$, as we anticipate that it could be of great importance for the forthcoming quark tomography at the EIC with jets~\cite{Arratia:2019vju,Page:2019gbf}.  

\let\oldaddcontentsline\addcontentsline
\renewcommand{\addcontentsline}[3]{}
\begin{acknowledgments}
{\bf Acknowledgements.} We are grateful to Cyrille Marquet, Farid Salazar, Gregory Soyez, and Werner Vogelsang for inspiring discussions. We thank the France-Berkeley-Fund from University of California at Berkeley for support. 
The work of F.Y. is supported in part by the U.S. Department of Energy, Office of Science, Office of Nuclear Physics, under contract number DE-AC02-05CH11231. The work of A.H.M. is supported in part by the U.S. Department of Energy Grant \# DE-FG02-92ER40699. 
\end{acknowledgments}
\let\addcontentsline\oldaddcontentsline

\let\oldaddcontentsline\addcontentsline
\renewcommand{\addcontentsline}[3]{}
\bibliographystyle{apsrev4-1}
\bibliography{refs}
\let\addcontentsline\oldaddcontentsline

\clearpage
\appendix

\begin{widetext}

\let\oldaddcontentsline\addcontentsline
\renewcommand{\addcontentsline}[3]{}
\section{Supplemental material}
\let\addcontentsline\oldaddcontentsline

\tableofcontents
\vspace{10 mm}

In this Supplemental Material, we will discuss in more detail two topics that have been briefly addressed in the
text of the Letter and which, as explained there, are deeply related to each other: the definition of jets in DIS and the
TMD factorisation of the cross-section for SIDIS as computed in the colour dipole picture to next-to-leading order (NLO).
We will notably show that some of the standard jet definitions used in the literature are inconsistent with TMD 
factorisation at NLO. In practice, this means that the respective definitions typically fail to properly reconstruct the jet structure generated by the struck quark.

\section{1 Jet definitions and NLO clustering conditions}
\label{sec:1}

In this section, we briefly review various standard jet definitions which have been used in phenomenological studies of DIS~\cite{Newman:2013ada,ZEUS:2002nms,ZEUS:2006xvn, ZEUS:2010vyw,H1:2007xjj,H1:2016goa,Arratia:2019vju,Page:2019gbf,ZEUS:2023zie,Arratia:2020ssx}, with the purpose of comparing them with the new definition that we introduced in Eq.~(5) of the Letter. 
In the original literature, these definitions are either formulated in the Breit frame, or, in some rare occasions, they are
given Lorentz-invariant formulations. In what follows, we shall re-express them in the dipole frame, which will be
convenient for the NLO calculations to be presented in the next section.  We always consider massless particles for simplicity.

\subsection{1.1 Longitudinally invariant jet algorithms in hadron-hadron collisions}

Longitudinally invariant generalised-$k_t$ algorithms~\cite{Ellis:1993tq,Dokshitzer:1997in,Wobisch:1998wt,Cacciari:2008gp,Cacciari:2011ma} in inclusive mode~\footnote{We consider throughout this section the inclusive mode~\cite{Salam:2010nqg} where a jet $i$ such that $d_{iB}$ is minimal is declared as a final jet, which is then removed from the list of particles.} for hadron-hadron collisions are defined with the distance measure
\begin{align}
    d_{ij}=\textrm{min}(k_{i\perp}^{2k},k_{j\perp}^{2k})\frac{\Delta R_{ij}^2}{R^2}\,,\quad d_{iB}=k_{i\perp}^{2k}\label{eq:LI-antikt}
\end{align}
where $\Delta R_{ij}^2$ is the square distance between the two particles $i$ and $j$ in the rapidity-azimuth plane. When there are only two particles in the final state, as it is the case at NLO for the SIDIS process we are considering, two particles $i$ and $j$ are recombined within a same jet if $d_{ij}\le \textrm{min}(d_{iB},d_{jB})$, or, equivalently,
$\Delta R_{ij}^2\le R^2$. (In this situation, the power $k$ plays no role in the clustering condition, but one should keep in mind that when the final state is made of more than two particles, this power modifies the ordering variable of the clustering sequence~\footnote{If there are more than 2 particles in the final state, the clustering condition is more complicated because $d_{ij}$ must also be smaller than all the others $d_{i'j'}$ for all pairs.}.) This algorithm is longitudinally invariant, so it can be implemented either in the Breit frame or in the dipole frame 
--- it gives the same jets in both cases. For small $\Delta R_{ij}$, we have~\cite{Liu:2013hba,Liu:2022ijp}
\begin{align}
    \Delta R_{ij}^2\approx\frac{2k_i\cdot k_j}{k_{i\perp}k_{j\perp}}\label{eq:deltaRij-def}
\end{align}
so that the clustering condition $\Delta R_{ij}^2\le R^2$ when only two particles are produced in the final state is 
\begin{align}
\textrm{\textbf{Clustering condition A:}}\quad    \frac{M_{ij}^2}{z_iz_j P_\perp^2R^2}\le \frac{1}{z^2}\label{eq:cluster-kt}
\end{align}
where $z$ and $\Pt$ are the jet longitudinal momentum fraction and transverse momentum in the so-called E-recombination scheme~\cite{Catani:1993hr,Blazey:2000qt},
in which the jet four-momentum is given by the sum of the four-momenta of the two clustered particles. To go from Eq.\,\eqref{eq:deltaRij-def} to Eq.\,\eqref{eq:cluster-kt}, we have used $M_{ij}^2\equiv (k_i+k_j)^2=2k_i\cdot k_j$ for on-shell
massless particles together with
 the collinear approximation where $k_{i\perp}\sim (z_i/z) P_\perp$. We shall refer to Eq.\,\eqref{eq:cluster-kt} as the clustering condition A.

\subsection{1.2 Jet algorithms in DIS}

\subsubsection{1.2.1 JADE-like algorithm from \cite{Webber:1993bm}}

A JADE-like algorithm~\cite{JADE:1986kta,JADE:1988xlj} for DIS has been proposed by Webber in \cite{Webber:1993bm}. It relies on the measure
\begin{align}
    d_{ij}=\frac{M_{ij}^2}{Q^2 R^2}\,,\quad d_{iB}= 2 x_{\rm Bj} \frac{k_i\cdot P}{Q^2}\,,\label{eq:dij-JadeWebber}
\end{align}
where the jet resolution parameter $d_{\rm cut}$ is written as $d_{\rm cut}=R^2$ following~\cite{Webber:1993bm}.
One advantage of this JADE-like definition is that the distance measure is Lorentz-invariant. At NLO, two particles $i$ and $j$ are recombined if $d_{ij}\le \textrm{min}(d_{iB},d_{jB})$. 
In the dipole frame, where $k_i\cdot P=z_i(q\cdot P)$ and hence  $d_{iB}=z_i$, this condition amounts to 
\begin{align}
    \frac{M_{ij}^2}{Q^2R^2}\le \textrm{min}(z_i,z_j)
\end{align}
We shall not consider this clustering condition in our NLO calculation, as the function $\textrm{min}(z_i,z_j)$ in the right hand side makes the computation more cumbersome, but we have checked that it gives the same NLO results modulo pure $\mathcal{O}(\alpha_s)$ corrections. In that sense, it is equivalent (at least to up to the NLO) to the jet definition with $p=1$ proposed in
Eq.~(5) in the main text. Note that the special choice $d_{iB}=z_i$ for the beam distance is crucial in that sense:  if it were fixed to $1$ as in the JADE algorithm for $e^+e^-$ annihilation, the distance measure would be identical to Eq.\,(5) in the main text with $p=0$, which violates TMD factorisation as shown in the main text.

\subsubsection{1.2.2 $k_t$-like algorithm from \cite{Catani:1992zp}}

In \cite{Catani:1992zp}, Catani, Dokshitzer and Webber extended the inclusive $k_t$ algorithm initially defined in $e^+e^-$ annihilation to the DIS case, using the distance measure
\begin{align}
    d_{ij}=2\textrm{min}(E_i^{2},E_j^{2})\frac{1-\cos\theta_{ij}}{R^2}\,,\quad d_{iB}=2E_i^2(1-\cos\theta_i)\label{eq:dij-kt}
\end{align}
computed in the Breit frame~\footnote{We recall that the Breit frame is the Lorentz frame in which the virtual photon is
a standing wave along the longitudinal ($z$) direction: $q^\mu=\delta^{\mu z}Q$ or, in LC notations, $q^+=-q^-=Q/\sqrt{2}$.}. It is clear that this distance measure is not boost invariant along the $\gamma^*A$ direction. In order to compute $d_{ij}$ and $d_{iB}$ in the dipole frame, we write these two quantities in terms of Lorentz invariant quantities, using 
\begin{equation}
    E_i=\frac{k_i\cdot\hat n}{Q}\label{eq:Breitframe-energy}
\end{equation}
with the 4-vector $\hat n\equiv 2x_{\rm Bj}P+q$ which reduces
to $\hat n = (Q, 0,0,0)$ in Breit frame and 
obeys $\hat n^2=Q^2$. Using $M_{ij}^2=2E_iE_j(1-\cos \theta_{ij})$, we have
\begin{align}
    d_{ij}=\frac{M_{ij}^2}{R^2}\textrm{min}\left(\frac{k_i\cdot \hat n}{k_j\cdot \hat n},\frac{k_j\cdot \hat n}{k_i\cdot \hat n}\right)\,,\quad d_{iB}=\frac{2(k_i\cdot \hat n)(k_i\cdot P)}{P\cdot\hat n}
\end{align}
In the dipole frame, these relations yield  
\begin{align}
    k_i\cdot \hat n
    &=\frac{z_iQ^2}{2}\left[1+\frac{\boldsymbol{k}_{i\perp}^2}{z_i^2Q^2}\right]\,,\quad d_{iB}=z_i^2Q^2\left[1+\frac{\boldsymbol{k}_{i\perp}^2}{z_i^2Q^2}\right]
\end{align}
Let us now consider the NLO clustering condition $d_{ij}<\textrm{min}(d_{iB},d_{jB})$ in the collinear limit where one can approximate $\boldsymbol{k}_{i\perp}\approx ({z_i}/{z})\Pt$. (Like before,  $\Pt$ and $z$ are the jet transverse momentum and longitudinal momentum fraction in the E-scheme.) A straightforward calculation gives the following clustering condition
in the dipole frame:
\begin{align}
    \frac{M_{ij}^2}{z_iz_jQ^2R^2}\left[1+\frac{ P_\perp^2}{z^2Q^2}\right]^{-1}\le 1\label{eq:kt-alg-dis}
\end{align}
up to corrections suppressed by powers of $R^2$.
This condition can be further simplified in the regime $Q^2\gg P_\perp^2$ dominated by aligned jet configuration where either $1-z\sim  P_\perp^2/Q^2$ or $z\sim  P_\perp^2/Q^2$. To leading power in the ratio $ P_\perp^2/Q^2$, we can replace Eq.\,\eqref{eq:kt-alg-dis} by
\begin{align}
     \frac{M_{ij}^2}{z_iz_jQ^2R^2}\le \frac{1}{z}\,.\label{eq:CDW}
\end{align}

\subsubsection{1.2.3 $e^+e^-$ generalised-$k_t$ algorithms applied to DIS in the Breit frame}

Another common choice of jet definition in the TMD literature is to use the generalised $k_t$ algorithms originally designed for $e^+e^-$ annihilation, which can be extended to DIS in the Breit frame (see e.g.~\cite{Gutierrez-Reyes:2018qez,Gutierrez-Reyes:2019msa,Gutierrez-Reyes:2019vbx}). Inclusive generalised-$k_t$ algorithms in DIS are defined with the spherically invariant distance measure~\cite{Cacciari:2011ma}
\begin{align}
    d_{ij}=\textrm{min}(E_i^{2k},E_j^{2k})\frac{1-\cos \theta_{ij}}{1-\cos R}\,,\quad d_{iB}=E_i^{2k}\label{eq:dij-antikt}
\end{align}
where energies and angles are defined in the Breit frame. Note that the distance measures are also not invariant by longitudinal boost.  As in the previous case (recall
Eq.\,\eqref{eq:Breitframe-energy}), the distance measure can be written with Lorentz invariant quantities as
\begin{align}
    d_{ij}=\frac{Q^2}{(k_i\cdot\hat n)(k_j\cdot\hat n)}\frac{M_{ij}^2}{2(1-\cos R)}\textrm{min}\left(\frac{(k_i\cdot \hat n)^{2k}}{Q^{2k}},\frac{(k_j\cdot \hat n)^{2k}}{Q^{2k}}\right)\,,\quad d_{iB}=\frac{(k_i\cdot \hat n)^{2k}}{Q^{2k}}
\end{align}
The condition $d_{ij}\le \textrm{min}(d_{iB},d_{jB})$ for two particles to be clustered within the same jet then reads
\begin{align}
    \frac{4\boldsymbol{k}_{ij\perp}^2}{z_i^2z_j^2Q^2}\left[1+\frac{\boldsymbol{k}_{i\perp}^2}{z_i^2Q^2}\right]^{-1}\left[1+\frac{\boldsymbol{k}_{j\perp}^2}{z_j^2Q^2}\right]^{-1}\le2( 1-\cos R)\,,
\end{align}
in the dipole frame.
In the collinear limit, one can again approximate $\boldsymbol{k}_{i\perp}\approx \frac{z_i}{z}\Pt$ giving the condition
\begin{align}
    \frac{4M_{ij}^2}{z_iz_j Q^2}\left[1+\frac{ P_\perp^2}{z^2Q^2}\right]^{-2}\le R^2\label{eq:dis-kt-DF}
\end{align}
up to power of $R^2$ corrections.
Let us consider the two regimes which contribute to leading twist (i) $1-z\sim  P_\perp^2/Q^2\ll 1$, (ii) $z\sim  P_\perp^2/Q^2\ll 1$. In the first case, we can approximate $1+ P_\perp^2/(z^2Q^2)\approx 1\approx z^{-1}$, while in the case (ii), we have $1+ P_\perp^2/(z^2Q^2)\approx Q^2/ P_\perp^2\approx z^{-1}$. Thus, to leading power in $ P_\perp^2/Q^2$, the clustering condition  Eq.\,\eqref{eq:dis-kt-DF} can be written 
\begin{align}
    \frac{4M_{ij}^2}{z_iz_jQ^2R^2}\le \frac{1}{z^2}\,.\label{eq:ee}
\end{align}
This differs from \eqn{eq:CDW} in the power of $1/z$ in the r.h.s. (together with  an irrelevant numerical factor which can be absorbed into a redefinition of $R$). It furthermore differs from the clustering condition A given by Eq.\,\eqref{eq:cluster-kt} in
the replacement $Q^2\leftrightarrow  P_\perp^2$. 

\subsubsection{1.2.4 New jet distance measure given by Eq.\,(5) in the main text}

We recall here that we propose in the main text the distance measure 
\begin{align}
    d_{ij}=\textrm{min}(k_{i\perp}^{2k},k_{j\perp}^{2k})\frac{M_{ij}^2}{z_iz_j Q^2 R^2}\,,\quad d_{iB}=k_{i\perp}^{2k}\label{eq:dij-new}
\end{align}
which is longitudinally boost invariant (we fix the parameter $p$ to 1 in Eq.\,(5) of the Letter as we already know from the discussion in the main text that it is the only consistent choice at NLO).
 
In order to simultaneously account for \texttt{(i)} the $k_t$ algorithm by Catani, Dokshitzer and Webber, \texttt{(ii)} the $e^+e^-$ generalised-$k_t$ algorithm used in the DIS  Breit frame, and \texttt{(iii)} our new jet distance measure, we shall use the generalised NLO clustering condition 
\begin{equation}
\textrm{\textbf{Clustering condition B:}}\quad      \frac{M_{ij}^2}{z_iz_j Q^2R^2}\le z^{-\beta}\label{eq:cluster-DIS}
\end{equation}
As demonstrated above, $\beta=1$ corresponds to the DIS $k_t$ algorithm defined in~\cite{Catani:1992zp}, cf. \eqn{eq:CDW},
 $\beta=2$ mimics the $e^+e^-$ generalised-$k_t$ algorithms applied to DIS in the Breit frame, cf. \eqn{eq:ee},
while $\beta=0$ is our new clustering condition, Eq.\,\eqref{eq:dij-new}. We shall refer to Eq.\,\eqref{eq:cluster-kt} as the clustering condition B. We will compute the SIDIS jet cross-section in the limit $Q^2\gg P_\perp^2$ for any value of $\beta$, although the choice $\beta=0$ is the only physical one, as we will argue in the next section.

\section{2 NLO SIDIS cross-section in the limit $Q^2\gg P_\perp^2\gg Q_s^2$ from the CGC}

In this section, we will study the virtual NLO corrections to the cross-section for jet production in SIDIS, with three main
objectives in mind: \texttt{(i)} to illustrate the (strong) dependence of the results upon the choice of a jet definition, as
summarised in the previous section, \texttt{(ii)} to demonstrate that the only jet definition which is consistent with
TMD factorisation at NLO is the condition (5) in the Letter with the choice $p=1$ (equivalently, this is the clustering
condition B, \eqn{eq:cluster-DIS} in the present material, with the choice $\beta=0$), and  \texttt{(ii)} to establish the
result for the virtual corrections shown in Eq.\,(12) in the Letter, which indeed corresponds to the ``physical'' jet definition
mentioned at point  \texttt{(ii)} above. We recall that the compatibility between the jet definition and TMD factorisation is
crucial in order to guarantee that the jet structure which is experimentally measured  faithfully reflect the partonic 
picture of the QCD process in perturbation theory, including the natural parton virtualities.

Before we proceed, it is important to stress that, from a diagrammatic point of view,  the ``virtual corrections'' of interest for us here include not only the {\it genuine} virtual diagrams --- the one-loop corrections to the amplitude ---, but also pieces of the would-be real corrections --- the NLO amplitudes involving three partons (a quark, an antiquark and a gluon) in the final state --- in the kinematical regime where the gluon jet is not well separated from the quark jet (meaning that the quark and the gluon are combined within a same jet by the jet algorithm). This is important for the present purposes since such (would-be) real NLO corrections are the only ones to be sensitive to the actual jet definition. For the one-loop corrections (and to strict NLO accuracy), the struck quark emerges alone in the final state and is unambiguously identified as {\it the} jet by all the clustering conditions.

As implicit in the previous discussion, we shall consider the case where the jet measured in SIDIS is generated by the quark. Our calculation is inclusive in the quark longitudinal momentum fraction $z_1$: that is, we shall consider both the situation where the quark is fast ($z_1\simeq 1$) in the dipole frame (meaning that the quark is the struck fermion in the Breit frame) and the situation where the quark is slow $z_1\ll 1$ (so the struck fermion is the antiquark). At NLO, the measured jet can reduce to the quark alone, or it can also include a gluon, that was radiated by either the quark, or the antiquark. Before we turn to the NLO corrections, we first review in the next subsection the leading order (LO) calculation of the SIDIS cross-section in the CGC/dipole picture and its leading twist approximation at $Q^2\gg P_\perp^2$, in which one recovers the TMD factorisation shown in Eq.\,(1) in the Letter.

\subsection{2.1 Brief review of the LO cross-section}

The LO SIDIS cross-section in the Colour Glass Condensate effective theory  reads~(see e.g.\,\cite{Iancu:2020jch,Caucal:2024cdq})
\begin{align}
 \left.\frac{\der \sigma_{\rm CGC}^{\gamma_{\rm T}^{\star}+A\to q+X}}{  \der^2\bt\der^2 \Pt\der z}\right|_{\rm LO} &=\frac{\alpha_{\rm em}e_f^2N_c}{(2\pi)^4}\ 2[z^2+(1-z)^2]\int\der^2\rxyt\der^2\rxpyt e^{-i\Pt \cdot (\rxyt-\rxpyt)}\Xi_{\rm LO}(\xt,\yt,\xt')\nonumber\\
 &\times \bar Q^2 K_1(\bar{Q}r_{x'y})K_1(\bar{Q}r_{xy})\frac{\rxpyt\cdot\rxyt}{r_{x'y}r_{xy}}\,.\label{eq:LO-full}
\end{align}
where  $\bar Q^2=z(1-z)Q^2$ and we define $\boldsymbol{r}_{xy}=\boldsymbol{x}_\perp -\boldsymbol{y}_\perp$ for any two transverse coordinate vectors $\xt$ and $\yt$. In this expression, the quark is measured with transverse momentum $\Pt$ and longitudinal momentum fraction $z$.  Physically, $\xt$ and $\xt'$ represent the transverse coordinates 
of the quark in the direct amplitude and the complex conjugate amplitude, respectively, and $\bt=(\xt+\xt')/2$ is the quark
impact parameter. The antiquark is not measured, so its coordinate is the same, and equal to  $\yt$, on both sides of the cut. The alternative situation, where the antiquark is measured but the quark is not, gives an identical contribution and is taken into account by an overall factor 2. If one integrates over the $\Pt$ and $z$ (with $z\in[0,1]$), one recovers the fully inclusive DIS cross-section at small $x$  and per unit $\bt$ (see e.g.~\cite{Ducloue:2019jmy}). 
The CGC colour structure in \eqn{eq:LO-full} is defined as \begin{align}
     \Xi_{\rm LO}(\xt,\yt,\xt')&\equiv\left\langle 1-D_{yx'}-D_{xy}+D_{xx'} \right\rangle_x\\
&=\int\der^2\qt \left(e^{i\qt\cdot\rxyt}-1\right)\left(e^{-i\qt\cdot\rxpyt}-1\right)\mathcal{D}(x,\qt),
\end{align}
where the Wilson-line operator  $D_{xy}=\frac{1}{N_c}\Tr(V({\xt})V^\dagger({\yt}))$ represents the $S$-matrix for the
elastic scattering of a quark-antiquark colour dipole.
The CCG average $\left \langle ...\right\rangle _x$ should be performed at the scale $x=x_{\rm Bj}$ as explained in the main text of the letter. In the second line of the above expression, we have written the colour correlator in momentum space for later convenience; in particular,
\begin{align}
\mathcal{D}(x,\qt)\equiv\int\frac{\der^2\rt}{(2\pi)^2}e^{-i\qt\cdot\rt}\frac{1}{N_c}\left\langle \Tr(V(\bt+\rt/2)V^\dagger(\bt-\rt/2))\right\rangle_x\,,
\end{align}
where we keep the dependence upon the impact parameter $\bt$ implicit.
Using this momentum-space representation, one can see that the integrals over $\rxyt$ and $\rxpyt $ in  Eq.\,\eqref{eq:LO-full} 
factorise from each other; after also integrating over $z$,  one finds
\begin{align}
    &\left.\frac{\der \sigma_{\rm CGC}^{\gamma_{\rm T}^{\star}+A\to q+X}}{ \der^2\bt \der^2 \Pt}\right|_{\rm LO} =\frac{\alpha_{\rm em}e_f^2N_c}{\pi^2}\int\der^2\qt \mathcal{D}(x,\qt)\int_0^{1/2}\der z \ [z^2+(1-z)^2]\, \Hcal_{\rm LO}^i(\Pt,\qt,z)\Hcal_{\rm LO}^{i*}(\Pt,\qt,z)\label{eq:LO-full-MS}
\end{align}
with 
\begin{align}
\mathcal{H}_{\rm LO}^i &\equiv \int\frac{\der^2\rxpyt}{2\pi} \ \  e^{-i\Pt\cdot\rxpyt}\left(e^{i\qt\cdot\rxpyt}-1\right)\frac{\bar Q K_1(\bar Qr_{x'y})}{r_{x'y}}\rxpyt^i\\
&=\frac{-i(\Pt-\qt)^i}{(\Pt-\qt)^2+\bar Q^2}-\frac{-i\Pt^i}{ P_\perp^2+\bar Q^2}\label{eq:LO-hf}
\end{align}
Note that we have integrated $z$ up to $1/2$ and multiplied the result by 2 thanks to the $z\leftrightarrow 1-z$ symmetry of the integrant. The modulus squared of the LO hard factor is readily obtained as
\begin{align}
    \mathcal{H}_{\rm LO}^i\mathcal{H}_{\rm LO}^{i*}&=\frac{ P_\perp^2}{[ P_\perp^2+\bar Q^2]^2}+\frac{(\Pt-\qt)^2}{\left[(\qt-\Pt)^2+\bar Q^2\right]}-\frac{2\Pt\cdot(\Pt-\qt)}{[(\Pt-\qt)^2+\bar Q^2][ P_\perp^2+\bar Q^2]}
\end{align}

So far, Eq.\,\eqref{eq:LO-full-MS} is just a rewriting of Eq.\,\eqref{eq:LO-full}. At this point we consider its leading-twist (LT)
approximation for $Q^2\gg  P_\perp^2$. In this limit, the integral over $z$ is dominated by the lower end point $z=0$, and more
precisely by values $z\sim P_\perp^2/Q^2$. To the LT accuracy of interest, the various contributions 
can be estimated as~\footnote{A simple way to derive these asymptotic behaviours is to make the change of variable $u=zQ^2$ and then to take $Q\to\infty$ in the upper limit of the $u$ integral.}
\begin{align}
    \int_0^{1/2}\der z \frac{1}{[ P_\perp^2+zQ^2]^2}\simeq\frac{1}{ P_\perp^2 Q^2}\,,\quad
     \int_0^{1/2}\der z \frac{1}{[ P_\perp^2+zQ^2][(\Pt-\qt)^2+z Q^2]}&\simeq\frac{\ln\left( P_\perp^2/(\Pt-\qt)^2\right)}{\big[P_\perp^2-(\Pt-\qt)^2 \big]Q^2}\,,\label{eq:approxz}
\end{align}
thus finally yielding the following, TMD-factorised, result:
\begin{align}
        \left.\frac{\der \sigma^{\gamma_{\rm T}^{\star}+A\to q+X}}{  \der^2 \Pt }\right|_{\rm LO} &=\frac{8\pi^2\alpha_{\rm em}e_f^2}{Q^2} \times  x\mathcal{F}_q^{(0)}(x,\Pt)\left[1+\mathcal{O}\left(\frac{P_\perp^2}{Q^2}\right)\right]\label{eq:LO-TMD}
\end{align}
with  the (sea) quark TMD in momentum space defined as
\begin{align}
    x\mathcal{F}_q^{(0)}(x,\Pt)\equiv \frac{N_c}{\pi^2} \int\der^2\bt \int\frac{\der^2\qt}{(2\pi)^2} \ \mathcal{D}(x,\qt)\left[1-\frac{\Pt\cdot(\Pt-\qt)}{( P_\perp^2-(\Pt-\qt)^2)}\,\ln \frac{ P_\perp^2}{(\Pt-\qt)^2}\right]\label{eq:quarkTMD-def}
\end{align}
Eq.\,\eqref{eq:LO-TMD} agrees with the results from the literature~\cite{Marquet:2009ca}. As it should be clear from its above derivation, it applies for relatively hard DIS with $Q^2\gg  P_\perp^2,Q_s^2$
(the corrections to it are suppressed by powers of $ P_\perp^2/Q^2$ and $Q_s^2/Q^2$). However, the saturation ``higher-twists'',
which describe multiple scattering with $P_\perp \sim Q_s$ and are  proportional to powers of
 $Q_s^2/ P_\perp^2$, are resummed to all orders within the TMD. In particular, the typical values of $q_\perp$ are of order
 $Q_s$ since for much larger values  $q_\perp\gg Q_s$,  $\mathcal{D}(x,\qt)$ is rapidly decreasing, as $1/\qt^4$
 \cite{Iancu:2003xm}.
 In the "dilute" limit $ P_\perp^2\gg Q_s^2$, where one can also neglect  the saturation effects, we can simplify the expression of the quark TMD by expanding the quantity in the square bracket in Eq.\,\eqref{eq:quarkTMD-def} for $q_\perp\sim Q_s\ll P_\perp$:\begin{align}
        x\mathcal{F}_q^{(0)}(x,\Pt) &=\frac{\alpha_s}{2\pi^2}\frac{1}{ P_\perp^2} \frac{1}{3}\int_0^{ P_\perp^2}\der^2\qt \ x\mathcal{G}_D^{(0)}(x,\qt)\,,\quad x\mathcal{G}_D^{(0)}(x,\qt)=\frac{N_c \qt^2}{2\pi^2\alpha_s}\int\der^2\bt\mathcal{D}(x,\qt)\label{eq:seaTMD-dilute}
\end{align}
where we have recognised the so-called gluon dipole TMD $x\mathcal{G}_D^{(0)}(x,\qt)$~\cite{Dominguez:2011wm}.
The $1/3$ coefficient physically comes from the integral of the $g\to q\bar q$ splitting function. Indeed, Eq.\,\eqref{eq:seaTMD-dilute} can be equivalently rewritten, in the small $x$ regime, in term of the gluon PDF $xG(x, P_\perp^2)$:
\begin{align}
    x\mathcal{F}_q^{(0)}(x,\Pt)  =\frac{\alpha_s}{2\pi^2}\frac{1}{ P_\perp^2}\int_0^1\der \xi \ P_{qg}(\xi)\frac{x}{\xi}G\left(\frac{x}{\xi}, P_\perp^2\right)\,,\quad xG^{(0)}(x, P_\perp^2)=\int_0^{ P_\perp^2}\der^2\qt \  x\mathcal{G}_D^{(0)}(x,\qt).
\end{align}
At tree-level, $\mathcal{D}(x,\qt)$ has no $x$ dependence, so the integral over $\xi$ can be performed analytically to yield
\begin{align}
  \int_0^{1}\der \xi \ P_{qg}(\xi)=  \int_0^1\der \xi \  T_R\left[\xi^2+(1-\xi)^2\right]=\frac{1}{3}\,,
\end{align}
in agreement with Eq.\,\eqref{eq:seaTMD-dilute}.  The emergence of the DGLAP splitting function has a simple interpretation: in order to produce a (sea) $q\bar q$ pair with large transverse momentum $P_\perp,$ a small-$x$ gluon with relatively small transverse momentum $q_\perp\ll P_\perp$ from the target must undergo a hard splitting.

\subsection{2.2 NLO diagrams with poles in transverse dim.\,reg.}

The subsequent calculation of NLO corrections to SIDIS will heavily rely on previously published NLO
calculations for dijets (or dihadrons) in DIS \cite{Caucal:2021ent,Taels:2022tza,Bergabo:2022tcu,Iancu:2022gpw,Bergabo:2023wed}
and for jets (or hadrons) in SIDIS \cite{Bergabo:2022zhe,Caucal:2024cdq}, that we shall simply adapt to the kinematics and the jet definitions of relevance for us here. 
In particular, with respect to \cite{Caucal:2024cdq}, we re-derive the real contribution to the SIDIS cross-section using the clustering condition B (in \cite{Caucal:2024cdq}, we only used the clustering condition A) and we extract the leading-twist contribution in the limit $Q^2\gg P_\perp^2$.

\begin{figure}
    \centering
    \includegraphics[width=0.5\textwidth]{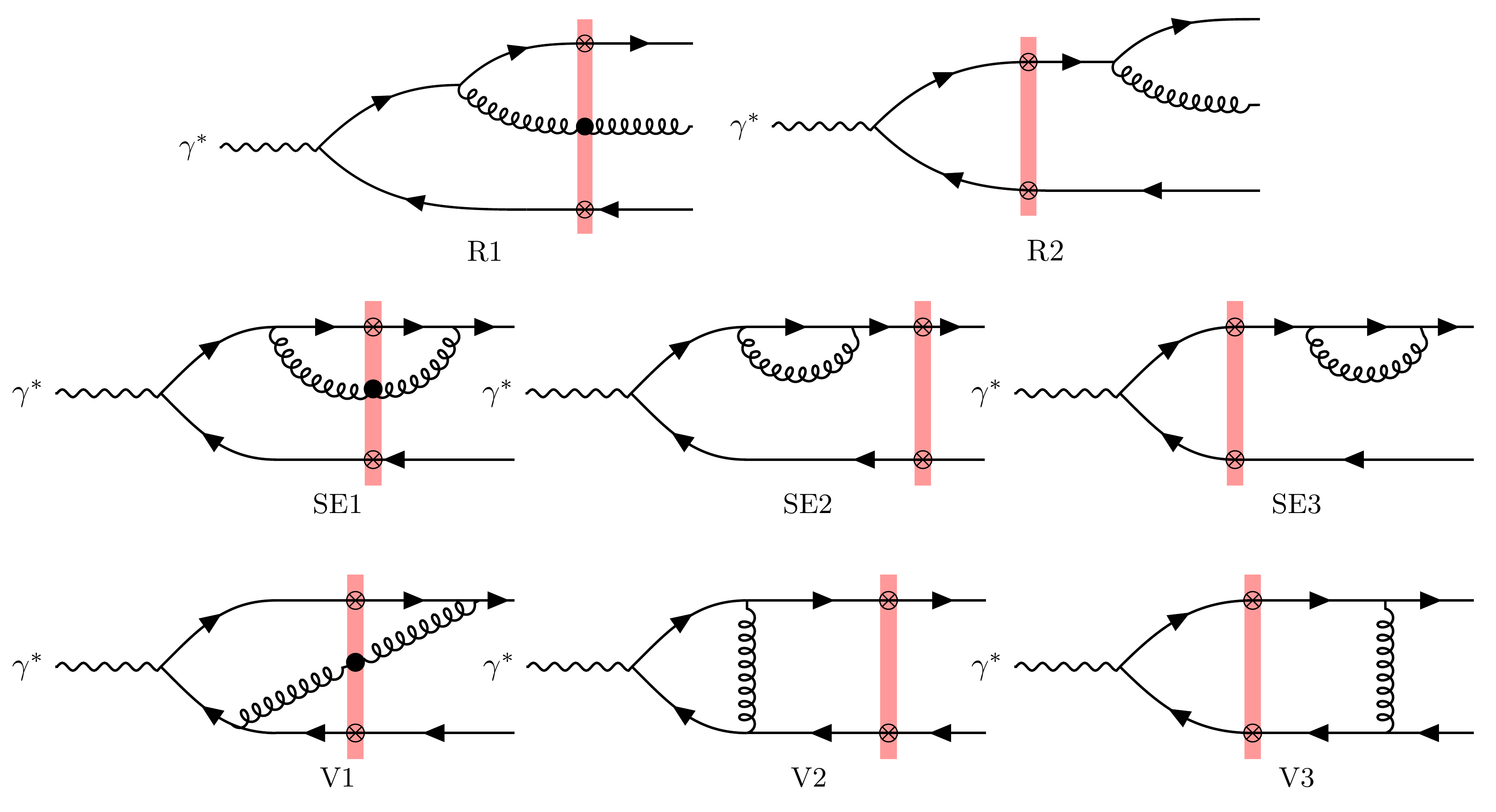}
    \caption{Feynman diagrams contributing to the NLO amplitude. The quark-antiquark exchanged diagrams, named with an additional line over the diagram's label, are not shown. Figure from~\cite{Caucal:2021ent}.}
    \label{fig:feynman-diagrams}
\end{figure}

The  Feynman graphs encoding the relevant NLO corrections are illustrated in Fig.~\ref{fig:feynman-diagrams} at the amplitude level. The two graphs in the first line describe
real gluon emissions, while all the other six graphs are virtual corrections
--- three quark self-energies and three vertex corrections in which the gluon is exchanged between the quark and the antiquark. We only display the real diagrams R1 and R2 where the gluon is emitted by the quark.  Clearly, there are corresponding graphs, to be denoted with a bar ($\overline{\rm R1}$, $\overline{\rm R2}$), in which the gluon is emitted by the antiquark.
Similarly, there are three antiquark self-energies ($\overline{\rm SE1}$, etc) and an additional vertex correction $\overline{\rm V1}$ where the gluon
is emitted by the quark prior to the scattering with the shockwave (SW) and absorbed by the antiquark after the SW. When computing the cross-section, the NLO real graphs are multiplied with each other, while the NLO virtual graphs are multiplied with the LO amplitude. For definiteness, we shall consider the situation where the jet measured by SIDIS has been initiated by the quark --- that is, the antiquark and the gluon are integrated over. This in particular implies that the gluon emissions by the antiquark in the final state (after crossing the SW) cannot contribute to the cross-section: their contributions must cancel between real and virtual corrections. In dimensional regularisation, where the final-state self-energies are identically zero (this applies to both  SE3 and  $\overline{\rm SE3}$), one also has  $\overline{\rm R2}\times\overline{\rm R2}^*=0$.

When using a jet definition which is infrared and collinear safe (IRC), the cross-section for SIDIS jets computed in
the CGC effective theory is free of  infrared and collinear divergences, by construction. Yet, such divergences can appear
in the individual diagrams and they are only guaranteed to cancel after all the diagrams are combined together. In this section, we display the results for the contributions which have both a $1/\varepsilon$ divergence in transverse dimensional regularisation and logarithmic divergences in the rapidity infrared cut-off $\Lambda^+$. We denote $z_0\equiv \Lambda^+/q^+$. 

We start with the $1/\varepsilon$ pole contributions generated by the virtual corrections.  They come from the diagrams $\rm SE2\times LO^*$ (and c.c.), $\rm V2\times LO^*$ (and c.c.), the ultraviolet (UV) singularity of the diagram $\rm SE1\times LO^*$ (and c.c.), and the UV singularity of the diagram $\overline{\rm R1}\times\overline{\rm R1}^*$. Regarding the latter, we classify it as a virtual correction (although the amplitude $\overline{\rm R1}$ describes a real gluon emission by the antiquark) because neither the gluon nor the antiquark are measured. The sum of these contributions is found as~\cite{Caucal:2024cdq} 
\begin{align}
    \left.\frac{\der\sigma_{\rm CGC}^{\gamma^*_{\rm T}+A\to q+X}}{\der^2\bt\der^2\Pt \der z}\right|_{V,\rm pole}&=\left.\frac{\der\sigma_{\rm CGC}^{\gamma^*_{\rm T}+A\to q+X}}{\der^2\bt\der^2\Pt\der z}\right|_{\rm LO}\times\frac{\alpha_sC_F}{\pi}\left\{\ln\left(\frac{z}{1-z}\right)\left[\frac{2}{\varepsilon}+\ln\left(e^{\gamma_E}\pi\mu^2r_{xy}r_{x'y}\right)\right]+\frac{1}{2}\ln^2\left(\frac{z}{1-z}\right)-\frac{\pi^2}{6}+\frac{5}{2}\right.\nonumber\\
    &\left.+\left(\ln\left(\frac{1-z}{z_0}\right)-\frac{3}{4}\right)\left(\frac{2}{\varepsilon}+\ln\left(e^{\gamma_E}\pi\mu^2\rxxtp^2\right)\right)-\frac{1}{4}\right\}\label{eq:virtual-pole}
\end{align}
with a slight abuse of notation since the quantity inside the curly bracket should appear inside the transverse coordinate integration in the LO cross-section given by Eq.\,\eqref{eq:LO-full}.

Among the other real diagrams, only the diagram $\rm R2\times R2^*$ --- real gluon emission by the measured quark
in the final state (i.e. after the collision with the SW) ---
displays a pole in dimensional regularisation coming from the collinear singularity. The calculation of this diagram therefore strongly depends on the jet definition one uses. One must distinguish two cases: \texttt{(i)} either the quark and the gluon are clustered inside the same jet --- the jet which is measured by SIDIS ---  with four-momentum given by the sum of the four-momenta of the quark and gluon; \texttt{(ii)}  or they are not clustered together, in which case they form two well separated jets, among which only one is measured and must be included in the SIDIS cross-section --- the other one must be integrated over. Clearly,
the distinction between gluon emissions inside and, respectively, outside the quark jet depends upon the jet definition, and here is where differences bewteen the various clustering criteria summarised in Sect.~1 start to be important.

As explained at length at the beginning of this section, we are primarily interested in the contributions of type \texttt{(i)},
i.e. in the situations where the gluon cannot be distinguished from the quark jet (``in-cone emissions''), which effectively provide virtual corrections to the cross-section for quark jet production. However, it turns out that the separation between in-cone and 
out-of-cone emissions introduces spurious rapidity divergences, which only cancel after the two types of contributions are added with each other. After this cancellation, we are left with logarithmically-enhanced contributions of the Sudakov type, which are important for us here. To correctly compute these logarithms, we shall include in our calculation the contributions of type \texttt{(ii)} (gluon emissions outside the quark jet) which are needed in order to cancel the spurious rapidity divergences. For that  purpose, it is sufficient to consider the situation where the quark jet is measured and the gluon jet is integrated out. The other case, where the gluon jet is measured and the quark one is integrated out, will be not considered.

The contribution to the cross-section from case  \texttt{(i)} (a gluon emitted {\it inside} the quark jet) can be computed as
\begin{align}
    \left.\frac{\der\sigma_{\rm CGC}^{\gamma^*_{\rm T}+A\to q+X}}{\der^2\bt\der^2\Pt}\right|_{\rm in}=\int_0^1\der z \ &\mu^\varepsilon\int\der^{2-\varepsilon}\kgt\int\der^2\ktone\int_{z_0}^1\der z_g\int_0^1\der z_1 \ \delta(z-(z_1+z_g))\delta(\Pt-(\ktone+\kgt))\nonumber\\
    &\times\left.\frac{\der\sigma_{\rm CGC}^{\gamma^*_{\rm T}+A\to qg+X}}{\der^2\bt\der^2\ktone\der^2\kgt\der z_1\der z_g}\right|_{\rm R2\times R2^*}\Theta_{\rm in}^{A/B}(\ktone,\kgt,z_1,z_g)\label{eq:incone-def}
\end{align}
where the step function $\Theta_{\rm in}^{A/B}$ enforces the clustering condition $A$ or $B$ respectively given by Eq.\,\eqref{eq:cluster-kt} and Eq.\,\eqref{eq:cluster-DIS}. In terms of the variables $\ktone$, $\kgt$, $z_1$ and $z_g$, they read
\begin{align}
    \textrm{Clustering condition A: }\,\quad \Theta_{\rm in}^A&=\Theta\left[\frac{z_1^2z_g^2 P_\perp^2R^2}{z^2}-(z_g\ktone-z_1\kgt)^2\right]\,,\label{eq:thetaA}\\
    \textrm{Clustering condition B: }\,\quad \Theta_{\rm in}^B&=\Theta\left[\frac{z_1^2z_g^2Q^2R^2}{z^\beta}-(z_g\ktone-z_1\kgt)^2\right].\label{eq:thetaB}
\end{align}
The real correction to the SIDIS cross-section coming from diagram $\rm R2\times R2^*$ is given by~\cite{Caucal:2021ent,Iancu:2022gpw,Bergabo:2023wed}
\begin{align}
    &\left.\frac{\der\sigma_{\rm CGC}^{\gamma^*_{\rm T}+A\to qg+X}}{\der^2\bt\der^2\ktone\der^2\kgt\der z_1\der z_g}\right|_{\rm R2\times R2^*}=\frac{\alpha_{\rm em}e_f^2N_c}{(2\pi)^6}\alpha_sC_F\int\der^2\rxyt\der^2\rxpyt \  \Xi_{\rm LO}(\xt,\yt,\xt')\frac{e^{-i(\kgt+\ktone)\cdot\rxxtp}}{(\kgt-\frac{z_g}{z_1}\ktone)^2}\nonumber\\
    &\times \left\{ 8\bar Q_{\rm R2}^2\left[(1-z_1-z_g)^2+(z_1+z_g)^2\right]\left(1+\frac{z_g}{z_1}+\frac{z_g^2}{2z_1^2}\right)\frac{\rxyt\cdot\rxpyt}{r_{xy}r_{x'y}}K_1(\bar Q_{\mathrm{R}2}r_{xy})K_1(\bar Q_{\mathrm{R}2}r_{x'y})\right\}\,,
    \label{eq:dijet-NLO-trans-R2R2-final}
\end{align}
with $\bar Q_{\rm R2}^2=(1-z_1-z_g)(z_1+z_g)Q^2$. This expression exhibits a collinear singularity at $\kgt= ({z_g}/{z_1})\ktone$. 

When the above expression is inserted in the integral
in Eq.\,\eqref{eq:incone-def}, the collinear singularity becomes a pole in dimensional regularisation. Moreover, the presence
of the step function $\Theta_{\rm in}^{A/B}$ introduces a double logarithmic divergence in the rapidity cut-off $z_0$. The latter comes from the phase-space where both $z_g$ and $\kgt$ are small, so
it cancels against the contribution associated with the case 
 \texttt{(ii)} where the quark and gluon form two separated jets and the gluon is integrated out  \cite{Taels:2022tza}.
 This contribution  (gluon emission {\it outside} the measured quark  jet) reads 
\begin{align}
     \left.\frac{\der\sigma_{\rm CGC}^{\gamma^*_{\rm T}+A\to q+X}}{\der^2\bt\der^2\Pt}\right|_{\rm out}=\int_0^1\der z \ &\int\der^{2}\kgt\int\der^2\ktone\int_{z_0}^1\der z_g\int_0^1\der z_1 \ \delta(z-z_1)\delta(\Pt-\ktone)\nonumber\\
    &\times\left.\frac{\der\sigma_{\rm CGC}^{\gamma^*_{\rm T}+A\to qg+X}}{\der^2\bt\der^2\ktone\der^2\kgt\der z_1\der z_g}\right|_{\rm R2\times R2^*}\left(1-\Theta_{\rm in}^{A/B}(\ktone,\kgt,z_1,z_g)\right)\,.\label{eq:outcone-def}
\end{align}

After combining the real corrections --- the "in" piece in Eq.\,\eqref{eq:incone-def} and the "out" piece in Eq.\,\eqref{eq:outcone-def} --- with the virtual corrections in Eq.\,\eqref{eq:virtual-pole} and by following the same steps as in e.g.~\cite{Caucal:2021ent,Taels:2022tza,Caucal:2022ulg} we find
(with  the same slight abuse of notation as in \eqn{eq:virtual-pole})
\begin{align}
    \left.\frac{\der\sigma_{\rm CGC}^{\gamma^*_{\rm T}+A\to q+X}}{\der^2\bt\der^2\Pt }\right|_{B\rm -def}&=\int_0^1\der z\left.\frac{\der\sigma_{\rm CGC}^{\gamma^*_{\rm T}+A\to q+X}}{\der^2\bt\der^2\Pt\der z}\right|_{\rm LO}\times\frac{\alpha_sC_F}{\pi}\left\{-\frac{1}{2}\ln^2(1-z)+\frac{3-2\beta}{2}\ln^2(z)-\frac{3(1-\beta/2)}{2}\ln(z)\right.\nonumber\\
        &-\frac{3}{4}\ln\left(\frac{Q^2r_{xx'}^2 R^2}{c_0^2}\right)+\frac{11}{2}-\frac{\pi^2}{2}+\ln\left(\frac{z}{1-z}\right)\ln\left(\frac{Q^2r_{xy}r_{x'y}R^2}{c_0^2}\right)-(1-\beta)\ln(z)\ln(1-z)\nonumber\\
        &\left.+\int_0^{1-z}\frac{\der z_g}{z_g}\ln\left(\frac{Q^2r_{xx'}^2R^2z_g^2}{c_0^2 z^{\beta}}\right)\left[1-J_0\left(\frac{z_g P_\perp r_{xx'}}{z}\right)\right]+\mathcal{O}(z)+\mathcal{O}(1-z)+\mathcal{O}(R^2)\right\}\label{eq:IRCsafe-Bdef}
\end{align}
for the clustering condition $B$ which depends on the parameter $\beta$ (see Eq.\,\eqref{eq:cluster-DIS}). As usual in this context, we include  the constant $c_0=2e^{-\gamma_E}$ with $\gamma_E$ the Euler-Mascheroni number in the arguments of the logarithms. In this expression, we have discarded powers of $z$ and $1-z$ inside the $z$ integration as they give contributions  suppressed by
powers of $ P_\perp^2/Q^2$ after performing the integral over $z$. On the other hand, for the clustering condition $A$, we have
\begin{align}
    &\left.\frac{\der\sigma_{\rm CGC}^{\gamma^*_{\rm T}+A\to q+X}}{\der^2\bt\der^2\Pt}\right|_{A\rm -def}=\int_0^1\der z\left.\frac{\der\sigma_{\rm CGC}^{\gamma^*_{\rm T}+A\to q+X}}{\der^2\bt\der^2\Pt\der z}\right|_{\rm LO}\times\frac{\alpha_sC_F}{\pi}\left\{-\frac{1}{2}\ln^2\left(\frac{z}{1-z}\right)+\ln\left(\frac{z}{1-z}\right)\ln\left(\frac{ P_\perp^2r_{xy}r_{x'y}R^2}{c_0^2}\right)+7\right.\nonumber\\
        &\left.-\frac{2\pi^2}{3}-\frac{3}{4}\ln\left(\frac{4 P_\perp^2r_{xx'}^2 R^2}{c_0^2}\right)+\int_0^{1-z}\frac{\der z_g}{z_g}\ln\left(\frac{ P_\perp^2r_{xx'}^2R^2z_g^2}{c_0^2z^2}\right)\left[1-J_0\left(\frac{z_g P_\perp r_{xx'}}{z}\right)\right]+\mathcal{O}(z)+\mathcal{O}(1-z)+\mathcal{O}(R^2)\right\}\label{eq:IRCsafe-Adef}
\end{align}
Importantly, for both clustering conditions ($A$ or $B$), all soft and collinear divergences have cancelled in the sum of real and virtual corrections, as they should (since we employ IRC safe jet definitions): indeed,   there is no $1/\varepsilon$ pole in Eqs.~\eqref{eq:IRCsafe-Bdef}--\eqref{eq:IRCsafe-Adef}, nor $\ln^2 z_0$, or $\ln z_0$, divergences. 
However, the two clustering conditions give different Sudakov double and single logarithms, as we shall shortly see.

So far, our calculation was quite general, in the sense that Eqs.~\eqref{eq:IRCsafe-Bdef}--\eqref{eq:IRCsafe-Adef} holds for generic values of the transverse momentum $P_\perp$ of the measured jet. Remember however that {\it (a)} TMD factorisation is expected only to  leading order in the leading twist approximation at $P_\perp^2\ll Q^2$ and {\it (b)} to extract the (double and single) Sudakov logarithms it suffices to consider the dilute limit at $P_\perp^2\gg Q_s^2$. So, in what follows we shall simplify the above results via approximations which are appropriate in the range $Q^2\gg P_\perp^2\gg Q_s^2$. 

To extract the leading twist contributions of Eqs.\,\eqref{eq:IRCsafe-Bdef}--\eqref{eq:IRCsafe-Adef}, i.e. their dominant contributions in the expansion in powers of $P_\perp^2/Q^2$,
we follow the same steps as in the calculation of the LO cross-section in the previous subsection. One should note that some terms inside the curly bracket contribute either via the endpoint $z=1$, or via $z=0$ (more precisely,  $z\sim P_\perp^2/Q^2\ll 1$), or through both. For example, the term expressed as an integral over $z_g$ only contributes via the endpoint $z=0$, since it vanishes as $z\to 1$. In the limit $z\to 0$, one can further approximate its contribution as
\begin{align}
    \int_0^{1-z}\frac{\der z_g}{z_g}\ln\left(\frac{Q^2r_{xx'}^2R^2z_g^2}{c_0^2 z^{\beta}}\right)\left[1-J_0\left(\frac{z_g P_\perp r_{xx'}}{z}\right)\right]
    &\simeq  \int_{\frac{c_0z}{P_\perp r_{xx'}}}^{1}\frac{\der z_g}{z_g}\ln\left(\frac{Q^2r_{xx'}^2R^2z_g^2}{c_0^2 z^{\beta}}\right)\nonumber \\
    &=\frac{1}{2}\ln\left(\frac{ P_\perp^2r_{xx'}^2}{z^2c_0^2}\right)\left[\ln\left(\frac{Q^2r_{xx'}^2R^2z^{1-\beta}}{c_0^2}\right)-\frac{1}{2}\ln\left(\frac{ P_\perp^2r_{xx'}^2}{c_0^2}\right)\right]\label{eq:intzg}
\end{align}
up to corrections suppressed by powers of $z\sim P_\perp^2/Q^2$. Recall that $z$ is the longitudinal fraction of the measured jet, that is, either the quark alone ($z=z_1$), or the quark accompanied by the gluon ($z=z_1+z_g$). So, the fact that $z\ll 1$ means that the fermion struck by the photon is the unmeasured antiquark ($z_2\simeq 1$). This in turn implies that, in the Breit frame, the measured jet propagates in the fragmentation region of the nuclear target. This is interesting since, as we shall shortly see, the contribution in \eqn{eq:intzg} leads to a Sudakov double logarithm. The integral over $z_g$ in Eq.\,\eqref{eq:IRCsafe-Adef} can be similarly estimated: its result can be inferred from \eqn{eq:intzg} by replacing $Q^2\to P_\perp^2$ and $\beta\to 2$. 

Furthermore, in the dilute limit $P_\perp \gg Q_s$, one can approximate $r_{xx'}\sim r_{xy}\sim r_{x'y} \sim c_0/ P_\perp$ inside the logarithms in Eqs.~\eqref{eq:IRCsafe-Bdef}--\eqref{eq:IRCsafe-Adef}. 
To see this, recall the factorised structure of the LO cross-section, cf. 
\eqn{eq:LO-full-MS}:  when  $P_\perp \gg q_\perp\sim Q_s$, the values of $ r_{xy}$ and $ r_{x'y} $ are fixed by 
Fourier transforms like \eqn{eq:LO-hf} to values of order $1/P_\perp$; the same is true for 
$r_{xx'}=|\rxyt-\rxpyt|$, cf.  \eqn{eq:LO-full}. After also integrating over $z$, we find
\begin{align}
    \left.\frac{\der\sigma^{\gamma^*_{\rm T}+A\to q+X}}{\der^2\bt\der^2\Pt}\right|_{A\rm -def}&=\left.\frac{\der\sigma^{\gamma^*_{\rm T}+A\to q+X}}{\der^2\bt\der^2\Pt}\right|_{\rm LO}\times\frac{\alpha_sC_F}{\pi}\left[-\ln(R)\ln\left(\frac{Q^2}{ P_\perp^2}\right)-\frac{3}{2}\ln(2R)+7-\frac{2\pi^2}{3}\right]\label{eq:Adef-final}\\
        \left.\frac{\der\sigma^{\gamma^*_{\rm T}+A\to q+X}}{\der^2\bt\der^2\Pt}\right|_{B\rm -def}&=\left.\frac{\der\sigma^{\gamma^*_{\rm T}+A\to q+X}}{\der^2\bt\der^2\Pt}\right|_{\rm LO}\times\frac{\alpha_sC_F}{\pi}\left[\frac{1}{2}\ln^2\left(\frac{Q^2}{ P_\perp^2}\right)+\left(\ln(R)-\frac{3\beta}{8}\right)\ln\left(\frac{Q^2}{ P_\perp^2}\right)-\frac{3}{2}\ln(R)+\frac{11}{2}-\frac{\pi^2}{2}\right]\label{eq:Bdef-final}
\end{align}
up to corrections suppressed by powers of $ P_\perp^2/Q^2$ and $Q_s^2/ P_\perp^2$. 

By inspection of the above results, one can
immediately notice the crucial role of the jet definition for the value of the Sudakov logarithms in the NLO impact factor. For the clustering condition $A$, the sum of the diagrams with a $1/\varepsilon$ divergence does not generate any Sudakov double logarithm, in sharp contrast with the clustering condition $B$. Physically this difference can be understood as follows. 
The double
logarithmic contribution in Eq.\,\eqref{eq:Bdef-final} comes from real gluon emissions which remain inside the quark jet~\footnote{Indeed, one can easily check that the double logarithmic piece in the r.h.s. of \eqn{eq:Bdef-final} has been generated
by the first term in the r.h.s. \eqn{eq:intzg} (the second term there vanishes after identifying $r_{xx'}\simeq c_0/ P_\perp$).},
cf. \eqn{eq:incone-def}, and which are {\it relatively} hard:  the transverse momenta of the quark and the gluon
obey $Q^2 \gg k_{1\perp}^2 \simeq k_{g\perp}^2\gg P_\perp^2$ (with $\ktone+\kgt=\Pt$, of course). Such hard in-cone emissions
are allowed by the clustering condition $B$, but not also by condition $A$. To see this, notice that in the hard kinematics
at hand, the quark-gluon relative momentum is itself hard: $(z_g\ktone-z_1\kgt)^2\sim k_{1\perp}^2 \sim k_{g\perp}^2$. Accordingly, it is only the second $\theta$--function in Eqs.~\eqref{eq:thetaA}--\eqref{eq:thetaB}, namely
$\Theta_{\rm in}^B$, which has a non-zero support in this kinematic range.
The difference between the two above results is indeed crucial for what follows. All the other NLO corrections that we
shall compute in what follows turn out to be independent of the clustering condition, so the dissymmetry  between 
Eq.\,\eqref{eq:Adef-final} and, respectively, Eq.\,\eqref{eq:Bdef-final} will transmit to the final results. The above argument
also suggests that the double logarithmic correction visible in \eqn{eq:Bdef-final} should have no physical consequences: this is associated with in-cone gluon emissions, which do not modify the final state. And indeed, we shall later find that this contribution is compensated by virtual emissions with similar kinematics. But the effect of the latter is independent of the clustering condition, as just mentioned. Since the (physically expected) real versus virtual compensation occurs for the clustering condition $B$ alone, this discussion already suggests that the clustering condition $A$ leads to unphysical features. We shall return to this point in our summary in Sect. 2.5.

 Even for the case of the clustering condition $B$, we observe a dependence upon the parameter $\beta$ entering at single logarithmic accuracy. After combining Eq.\,\eqref{eq:Bdef-final} with the remaining diagrams (those which are finite when $\varepsilon\to 0$), we will see that only the choice $\beta=0$ in the clustering condition $B$ gives a result which is consistent with TMD factorisation.

\subsection{2.3 Virtual diagrams without poles}

We now consider the NLO virtual graphs which are finite as $\varepsilon\to 0$. The contribution to the SIDIS jet cross-section coming from these diagrams does not depend on the jet definition since only one particle is produced in the final state, necessarily giving a single jet. We detail the extraction of the leading twist contribution in the limit $Q^2\gg  P_\perp^2$ for the case of the
``dressed'' self-energy correction SE1 in Fig.~\ref{fig:feynman-diagrams}, where the gluon crossies the shock-wave (hence its propagator is dressed with a Wilson line).
The corresponding results for the vertex correction $\rm V1\times LO^*$ can be similarly obtained and will be presented in section 2.3.2. The other virtual graphs displayed in Fig.~\ref{fig:feynman-diagrams} either vanish in dimensional regularisation as diagram $\rm SE3$ or identically cancel against real corrections (we have $\rm\overline{SE1}\times LO^*+\overline{\rm R1}\times \overline{\rm R2}^*=0$ or $\rm\overline{V1}\times LO^*+\rm R1\times \overline{\rm R2}^*=0$)~\cite{Caucal:2024cdq,Bergabo:2022zhe}.


\subsubsection{2.3.1 Dressed self-energy correction on the quark}

Consider the self-energy correction  $\rm SE1\times LO^*$ to the SIDIS cross-section.
The UV-divergent piece (a pole in $1/\varepsilon$) of this contribution has already  been included in Eq.\,\eqref{eq:virtual-pole}.
After subtracting this piece and also the instantaneous term (which turns out to be power suppressed in the limit $Q^2\gg P_\perp^2$), the remaining, finite, contribution to the SIDIS cross-section reads~\cite{Caucal:2021ent,Caucal:2024cdq}
\begin{align}
 \left.\frac{\der \sigma_{\rm CGC}^{\gamma_{\rm T}^{\star}+A\to q+X}}{  \der^2\bt\der^2 \Pt \der z}\right|_{\rm SE1\times LO^*} &=\frac{\alpha_{\rm em}e_f^2N_c}{(2\pi)^4} \int\der^2\rxyt\der^2\rxpyt\frac{\der^2\rzxt}{\rzxt^2} e^{-i\Pt \cdot \rxxtp}\frac{\alpha_s}{\pi^2}2z(z^2+(1-z)^2)\frac{\bar Q K_1(\bar{Q}r_{x'y})\rxpyt^i}{r_{x'y}}  \nonumber\\
 &\times \int_0^{z} \frac{\der z_g}{z_g}\Bigg\{\left(1-\frac{z_g}{z}+\frac{z_g^2}{2z^2}\right)\left[ \bar QK_1(QX_V)\frac{\sqrt{1-z}\RtS^i}{X_V} e^{-i\frac{z_g}{z}\Pt\cdot\rzxt}\Xi_{\rm NLO,1}(\xt,\yt,\zt,\xt')\right.\nonumber\\
 &\left.-\bar QK_1(\bar{Q}r_{xy})\exp\left(-\frac{\rzxt^2}{\rxyt^2e^{\gamma_E}}\right)\frac{\rxyt^i}{r_{xy}}C_F \Xi_{\rm LO}(\xt,\yt,\xt')\right]\Bigg\}_{z_\star}\,,\label{eq:SE1-full}
\end{align}
where we recall that $\bar Q^2=z(1-z) Q^2$ and the additional notations read as follows
\begin{align}
    X_V^2&= (1-z)(z-z_g)\rxyt^2 + z
    _g(z-z_g)\rzxt^2 +(1-z)z_g\rzyt^2\,, \label{eq:XVdef}\\
    \RtS&=\rxyt+\frac{z_g}{z}\rzxt,\\
\Xi_{\rm NLO,1}(\xt,\yt,\zt,\xt')&=\frac{N_c}{2}\left\langle 1-D_{yx'}-D_{xz}D_{zy}+D_{zx'}D_{xz} \right\rangle_x- \frac{1}{2N_c}\left\langle 1-D_{xy}-D_{yx'}+D_{xx'} \right\rangle_x.
\end{align}
Notice the subscript $z_\star$ on the accolades in \eqn{eq:SE1-full}: $\{\cdots\}_{z_\star}$. This notation expresses the fact that
the rapidity divergence as $z_g\to 0$ has already been subtracted off, using an arbitrary "rapidity factorisation scale" $z_\star=k_\star^+/q^+$. More precisely, our notation means
\begin{equation}
    \int_0^{a}\frac{\der z_g}{z_g}\left\{f(z_g)\right\}_{z_\star}\equiv \int_0^a\frac{\der z_g}{z_g}\left[f(z_g)-\Theta(z_\star-z_g)f(0)\right] \,,
\end{equation}
for any function $f(z_g)$ which has a well-defined limit as $z_g\to 0$. As discussed in the main text of the letter, we use $z_\star=x_\star  P_\perp^2/Q^2$ in order to separate the NLO impact factor from the (kinematically constrained) BK evolution of the dipole cross-section. The dependence upon the arbitrary $x_\star$ number can be used to gauge the sensitivity of the final result to the arbitrary separation between high energy and DGLAP+Sudakov evolution.

Here comes an important observation: the leading twist piece of the expression Eq.\,\eqref{eq:SE1-full} can be obtained from the transverse coordinate domain where $r_{zx}\sim 1/Q \ll r_{xy}\sim 1/P_\perp$.  Physically, this means that the virtual
quark-gluon fluctuation is much harder than the measured quark.
Similar to the LO calculation, we shall systematically consider the aligned jet configuration such that $1-z\sim  P_\perp^2/Q^2\ll 1$ (the other endpoint $z\to 0$ is power suppressed). In this limit, the colour structure simply reduces to the LO one times a Casimir factor $C_F$, namely
\begin{align}
\Xi_{\rm NLO,1}(\xt,\yt,\zt,\xt')&= C_F \left\langle 1-D_{yx'}-D_{xy}+D_{xx'} \right\rangle_x+\mathcal{O}(Q_s^2\rzxt^2)\\
&\simeq C_F\int\der^2\qt \left(e^{i\qt\cdot\rxyt}-1\right)\left(e^{-i\qt\cdot\rxpyt}-1\right)\mathcal{D}(x,\qt)
\end{align}
In Eq.\,\eqref{eq:SE1-full}, one can also remove the phase $e^{-i\frac{z_g}{z}\Pt\cdot\rzxt}$ since the argument is of order $P_\perp/Q\ll 1$. Finally, the expression of $X_V$ in \eqn{eq:XVdef} simplifies as well:
in the limit we are considering, $r_{zx}\ll r_{xy}\simeq r_{zy}$, which gives
\begin{align}
X_V^2\simeq z(1-z)\rxyt^2+z_g(1-z_g)\rzxt^2\,,\qquad\RtS\simeq \rxyt.
\end{align}

After these simplifications, and integrating over the rapidity $z$ of the jet, Eq.\,\eqref{eq:SE1-full} reads~\cite{Caucal:2021ent,Caucal:2024cdq}
\begin{align}
 &\left.\frac{\der \sigma^{\gamma_{\rm T}^{\star}+A\to q+X}}{ \der^2\bt \der^2 \Pt}\right|_{\rm SE1} =\frac{\alpha_{\rm em}e_f^2N_c}{2\pi} \frac{4\alpha_s C_F}{\pi^2} \int\der^2\qt \mathcal{D}(x,\qt) \int_0^1\der z\int_{x_\star P_\perp^2/Q^2}^1\frac{\der z_g}{z_g}\left(1-z_g+\frac{z_g^2}{2}\right)\mathcal{H}_{\rm LO}^{i*} \mathcal{H}_{\rm SE1}^i.
\end{align}
As anticipated, when $Q^2\gg P_\perp^2$ the integral over $z$  is dominated by the endpoint at $z=1$. The factor of $4$ comes from the overall factor of $2$ in the first line of Eq.\,\eqref{eq:SE1-full} and a factor of 2 coming from the complex conjugate term not included in Eq.\,\eqref{eq:SE1-full}. The "hard factor" associated with the dressed self-energy $\rm SE1$ is defined as
\begin{align}
\mathcal{H}_{\rm SE1}^i&\equiv \int\frac{\der^2\rzxt}{(2\pi)}\int\frac{\der^2\rxyt}{(2\pi)}e^{-i\Pt\cdot\rxyt}\left(e^{i\qt\cdot\rxyt}-1\right)\frac{\rxyt^i}{\rzxt^2}\left[\frac{\bar Q K_1\left(\bar Q\sqrt{\rxyt^2+\omega\rzxt^2}\right)}{\sqrt{\rxyt^2+\omega\rzxt^2}}-\frac{\bar QK_1(\bar{Q}r_{xy})\exp\left(-\frac{\rzxt^2}{\rxyt^2e^{\gamma_E}}\right)}{r_{xy}}\right]\label{eq:HSE1}
\end{align}
with the new notation
\begin{align}
    \omega\equiv\frac{z_g(1-z_g)}{z(1-z)}\approx \frac{z_g(1-z_g)}{1-z}
\end{align}
The subtraction term within the square brackets, involving the Gaussian factor $\exp\left(-\frac{\rzxt^2}{\rxyt^2e^{\gamma_E}}\right)$, is a UV regulator of the $1/\rzxt^2$ singularity in diagram $\rm SE1$. This regulator is added to and subtracted from the total cross-section, in such a way that the overall result does not depend on it~\cite{Caucal:2021ent,Caucal:2024cdq}. The integrals over $\rzxt$ and $\rxyt$ can be exactly performed using the momentum space representation of the Bessel functions and yield the relatively simple result
\begin{align}
\mathcal{H}_{\rm SE1}^i&= \frac{-i(\Pt^i-\qt^i)}{2((\Pt-\qt)^2+\bar Q^2)}\left[\ln\left(\frac{1}{\omega}\right)-\frac{\bar Q^2}{(\Pt-\qt)^2}\ln\left(1+\frac{(\Pt-\qt)^2}{\bar Q^2}\right)\right]\nonumber\\
&+\frac{i\Pt^i}{2( P_\perp^2+\bar Q^2)}\left[\ln\left(\frac{1}{\omega}\right)-\frac{\bar Q^2}{ P_\perp^2}\ln\left(1+\frac{ P_\perp^2}{\bar Q^2}\right)\right]
\end{align}
Note that this hard factor is leading power, as one easily sees this by comparing with the LO one in Eq.\,\eqref{eq:LO-hf}
(recall that $\bar Q^2\sim P_\perp^2$ for the relevant value $1-z\sim P_\perp^2/Q^2$). The last steps consist in doing the contraction over the index $i$, expanding the result for $P_\perp \gg q_\perp$ and finally performing the integrals over $z_g$
and $z$. The integral over $z$ is performed along the lines of Eq.~\eqref{eq:approxz}
 and gives back the LO cross-section in the limit $ P_\perp^2\gg Q_s^2$. We thus get
\begin{align}
     \left.\frac{\der \sigma^{\gamma_{\rm T}^{\star}+A\to q+X}}{  \der^2 \Pt}\right|_{\rm SE1} =\left.\frac{\der \sigma^{\gamma_{\rm T}^{\star}+A\to q+X}}{  \der^2 \Pt}\right|_{\rm LO}\times\frac{\alpha_s C_F}{\pi}&\left[-\frac{3}{2}+\frac{\pi^2}{6}+\frac{3}{4}\ln\left(\frac{Q^2}{ P_\perp^2}\right)-\ln\left(\frac{Q^2}{x_\star  P_\perp^2}\right)\ln\left(\frac{Q^2}{ P_\perp^2}\right)+\frac{1}{2}\ln^2\left(\frac{Q^2}{x_\star  P_\perp^2}\right)\right.\nonumber\\
     &\left.-\frac{2}{3}\left(-\frac{3}{4}+\ln\left(\frac{Q^2}{x_\star  P_\perp^2}\right)\right)\right]\label{eq:SE1-final}
\end{align}
To summarise, the (UV-regularised) quark self-energy correction where the gluon crosses the shock-wave 
--- the amplitude  SE1 in Fig.~\ref{fig:feynman-diagrams} ---
contributes to the Sudakov double logarithm with a coefficient $-\frac{\alpha_sC_F}{2\pi}\ln^2(Q^2/ P_\perp^2)$. 
It is also interesting to notice that the double-logarithmic terms of the type
\beq
\ln\left(\frac{Q^2}{ P_\perp^2}\right)\,\ln\left(\frac{1}{x_*}\right), \label{eq:mix}
\eeq
which would mix the Sudakov logarithms with the logarithm of the arbitrary factorisation scale $x_*$, exactly cancel in the r.h.s. of 
\eqn{eq:SE1-final}. We shall comment on the meaning of this cancellation in the summary section 2.5.

\subsubsection{2.3.2 Dressed vertex correction from the antiquark to the quark}

We then consider the vertex correction denoted as V1 in in Fig.~\ref{fig:feynman-diagrams}: the gluon is first emitted by the antiquark, it crosses the shockwave background field and then it is absorbed by the quark.
 Its contribution to the SIDIS cross-section is~\cite{Caucal:2024cdq}
\begin{align}
    &\left.\frac{\der \sigma_{\rm CGC}^{\gamma_{\rm T}^{\star}+A\to q+X}}{  \der^2\bt\der^2 \Pt \der z}\right|_{\rm V1\times LO^*} =\frac{\alpha_{\rm em}e_f^2N_c}{(2\pi)^4} \int\der^2\rxyt\der^2\rxpyt\der^2\rzxt e^{-i\Pt \cdot \rxxtp}   \frac{\alpha_s}{\pi^2} 2z(1-z) Q\bar{Q} \frac{K_1(\bar{Q}r_{x'y})}{r_{x'y}} \Xi_{\rm NLO,1}(\xt,\yt,\zt,\xt')\nonumber\\
    & \times \int_0^{z}\frac{\der z_g}{z_g} \left\{ e^{-i\frac{z_g}{z}\kt\cdot\rzxt}\frac{K_1(QX_V)}{X_V} \left[- \frac{z_g(z-z_g)(1+z_g-2z)^2}{2z^2(1-z)}\frac{(\RtV\times\rxpyt)(\rzxt\times\rzyt)}{\rzxt^2\rzyt^2}  \right.\right.\nonumber\\
    &\left.\left.- [z(z-z_g)+(1-z)(1-z+z_g)]\left(1-\frac{z_g}{z} \right) \left(1+\frac{z_g}{1-z}\right) \left(1-\frac{z_g}{2z}-\frac{z_g}{2(1-z+z_g)}\right) \frac{(\RtV\cdot\rxpyt)(\rzxt\cdot\rzyt)}{\rzxt^2\rzyt^2} \right]\right\}_{z_\star}
\end{align}
where the power-suppressed instantaneous contribution has already been discarded.
We employ the same approximations as for the previous diagrams. Namely, for $z$ close to 1, $\RtV=\rxyt-\frac{z_g}{1-z+z_g}\rzyt\simeq -\rzxt$. In the same limit $1-z\sim P_\perp^2/Q^2\ll 1$, the prefactors depending upon $z$ and $z_g$ in front
of both terms inside the square bracket become equal to each other, namely they both reduce to
\begin{align}
    \frac{-z_g(1-z_g)^3}{2(1-z)}
\end{align}
so that we can combine together the two vectorial structures
\begin{align}
     \frac{(\RtV\times\rxpyt)(\rzxt\times \rzyt)+(\RtV\cdot\rxpyt)(\rzxt\cdot \rzyt)}{\rzxt^2\rzyt^2}&\approx- \frac{(\rzxt\times\rxpyt)(\rzxt\times \rxyt)+(\rzxt\cdot\rxpyt)(\rzxt\cdot \rxyt)}{\rzxt^2\rxyt^2}\\
     &=-\frac{\rxyt\cdot\rxpyt}{\rxyt^2}
\end{align}
In the end, we can cast this diagram (plus its complex conjugate) into the simpler form
\begin{align}\label{eq:V1}
 &\left.\frac{\der \sigma^{\gamma_{\rm T}^{\star}+A\to q+X}}{  \der^2\bt\der^2 \Pt}\right|_{\rm V1} =\frac{\alpha_{\rm em}e_f^2N_c}{2\pi} \frac{2\alpha_s C_F}{\pi^2}\int\der^2\qt \mathcal{D}(x,\qt)  \int_0^1\der z\int_{x_\star P_\perp^2/Q^2}^1\frac{\der z_g}{z_g}\frac{z_g(1-z_g)^3}{1-z}\mathcal{H}_{\rm LO}^{i*}\Hcal_{\rm V1}^i 
\end{align}
with the hard factors
\begin{align}
    \mathcal{H}_{\rm V1}^i&= \int\frac{\der^2\rzxt}{(2\pi)}\int\frac{\der^2\rxyt}{(2\pi)}e^{-i\kt\cdot\rxyt}\left(e^{i\qt\cdot\rxyt}-1\right)\frac{\rxyt^i}{\rxyt^2}\frac{\bar Q K_1\left(\bar Q\sqrt{\rxyt^2+\omega\rzxt^2}\right)}{\sqrt{\rxyt^2+\omega\rzxt^2}}\\
  &=\frac{(-i)}{2\omega}\left[\frac{\Pt^i-\qt^i}{(\Pt-\qt)^2}\ln\left(1+\frac{(\Pt-\qt)^2}{\bar Q^2}\right)-\frac{\Pt^i}{ P_\perp^2}\ln\left(1+\frac{ P_\perp^2}{\bar Q^2}\right)\right]
\end{align}
The overall factor $1/\omega\propto 1-z$ in the r.h.s. of the above expression for $\mathcal{H}_{\rm V1}^i$
compensates the would-be divergence at $z\to 1$ of the rational function in \eqn{eq:V1}.
We finally perform the integrals over $z$ and $z_g$ and take both the high-$Q^2$ limit $Q^2\gg P_\perp^2$ and the dilute limit $P_\perp\gg Q_s\sim q_\perp$ in order to simplify the calculation. This gives:
\begin{align}
     &\left.\frac{\der \sigma^{\gamma_{\rm T}^{\star}+A\to q+X}}{\der^2 \Pt}\right|_{\rm V1} =\left.\frac{\der \sigma^{\gamma_{\rm T}^{\star}+A\to q+X}}{\der^2 \Pt}\right|_{\rm LO}\times  \frac{\alpha_s C_F}{\pi}\frac{3}{2}\left(-\frac{3}{4}+\frac{1}{2}\ln\left(\frac{Q^2}{x_\star P_\perp^2}\right)\right)\label{eq:V1-final}
\end{align}

\subsection{2.4 Real diagrams without poles}

Besides the virtual graphs in Fig.\,\ref{fig:feynman-diagrams}, there are contributions to the SIDIS jet cross-section coming from real diagrams without collinear divergence where the quark jet is measured and the antiquark and the gluon are integrated out. Since these diagrams have no collinear divergences, the jet definition does not play any role in the "narrow jet" approximation $R\ll 1$. More formally, one can integrate over the full antiquark+gluon phase space, so that the resulting expression is in fact similar to the virtual corrections computed above. That is also why such diagrams actually contribute to the "virtual" part of the NLO correction to the quark TMD ; we shall indeed observe that the quark jet $\Pt$ in these contributions comes from the quark
TMD (``intrinsic $P_\perp$'') and not from the recoil associated with the gluon emission.

\subsubsection{2.4.1 Real gluon emission from the unmeasured antiquark before the shock-wave}

We start with the gluon emission by the antiquark before the SW both in the amplitude and in the complex conjugate amplitude. The  respective contribution to the cross-section, denoted as $\overline{\rm R1}\times\overline{\rm R1}^*$, is UV singular because of the $1/\rzyt^2$ (with $\rzyt=\zt-\yt$) kernel in the integrand. To regularise this divergence, a UV regulator is subtracted from the expression for  $\overline{\rm R1}\times\overline{\rm R1}^*$ and added to the pole term in
 Eq.\,\eqref{eq:virtual-pole} (cf. the discussion after \eqn{eq:HSE1}). 
 After including this regulator,    the diagram $\overline{\rm R1}\times \overline{\rm R1}^*$ reads
\begin{align}
    \left.\frac{\der \sigma_{\rm CGC}^{\gamma_{\rm T}^{\star}+A\to q+X}}{\der^2\bt \der^2 \Pt \der z}\right|_{\overline{\rm R1}\times \overline{\rm R1}^*} &=\frac{\alpha_{\rm em}e_f^2N_c}{(2\pi)^4}\int\der^2\rxyt\der^2\rxpyt\frac{\der^2\rzyt}{\rzyt^2} e^{-i\Pt \cdot \rxxtp} \frac{\alpha_s}{2\pi^2} z^2(\bar z^2+(1-z)^2)\left(z^2 + (1-z)^2\right)   \nonumber\\    &\times \int_0^{1-z} \frac{\der z_g}{z_g}\left\{\left[Q^2K_1(QX_R)K_1(QX_R') \frac{\RtRb\cdot\RtRb'}{X_R X_R'} \,\Xi_{\rm NLO,4}(\xt,\yt,\zt,\xt')\right. \right.\nonumber \\ 
    &\left.\left.  -  \frac{Q^2}{z(1-z)}K_1(\bar Q r_{xy})K_1(\bar Q r_{x'y}) e^{-\frac{\rzyt^2}{\rxxtp^2e^{\gamma_E}}}\frac{\rxyt\cdot\rxpyt}{r_{xy}r_{x'y}} \,C_F\Xi_{\rm LO}(\xt,\yt,\xt')\right]\right\}_{z_\star}\,,\label{eq:R1bar}
\end{align}
with $\RtR'=\rxpyt+\frac{z_g}{z+z_g}\rzxpt$, $\RtRb'=-\rxpyt+\frac{z_g}{1-z}\rzyt$, $X_R^2=z\bar z \rxyt^2+zz_g\rzxt^2+\bar z z_g\rzyt^2$ and $X_R'^2=z\bar z \rxpyt^2+zz_g\rzxpt^2+\bar z z_g\rzyt^2$, $\bar z=1-z-z_g$. Once again, we do not show the instantaneous piece as it is power suppressed.  The second term in the square bracket in Eq.\,\eqref{eq:R1bar} is meant to regulate this UV divergence, as just explained. The CGC correlator of this diagram is built from dipoles only,
\begin{align}
        \Xi_{\rm NLO,4}(\xt,\yt,\zt,\xt')&\equiv\frac{N_c}{2} \left\langle D_{xx'}-D_{xz}D_{zy}-D_{yz}D_{zx'}+1\right\rangle-\frac{1}{2N_c}\left\langle D_{xx'} - D_{xy} -  D_{yx'} + 1\right\rangle
\end{align}

Let now focus on the kinematic regime where the measured quark has $z\sim  P_\perp^2/Q^2$
and $|\zt-\yt|\sim 1/Q\ll |\xt-\yt|\sim 1/P_\perp$. We have in mind that $|\zt-\yt|$, physically conjugated to the transverse momentum $\kgt$ of the gluon, is controlled by sizes much smaller than $|\xt-\yt|=\mathcal{O}(1/P_\perp)$, i.e. $\kgt^2\gg  P_\perp^2$. It implies that $\RtRb\approx -\rxyt$. Note that the configuration where the measured quark has $1-z\sim  P_\perp^2/Q^2$ is power suppressed since the integral over $z_g$ vanishes as $z\to 1$.
In the limit $z\sim  P_\perp^2/Q^2\ll 1$ and $r_{zy}\ll r_{xy}$, the (squared) transverse distance $X_R$ in the argument of the Bessel function reduces to
\begin{align}
    X_R&=z(1-z-z_g)\rxyt^2+zz_g\rzxt^2+(1-z-z_g)z_g\rzyt^2\\
    &\approx z\rxyt^2 +z_g(1-z_g)\rzyt^2
\end{align}
We also systematically expand the rational function in $z$ and $z_g$ by keeping the leading power piece in $z$ and we expand the color correlator $\Xi_{\rm NLO,4}$ around $\zt=\yt$ such that
\begin{align}
        \Xi_{\rm NLO,4}(\xt,\yt,\zt,\xt')&\equiv\frac{N_c}{2} \left\langle D_{xx'}-D_{xz}D_{zy}-D_{yz}D_{zx'}+1\right\rangle-\frac{1}{2N_c}\left\langle D_{xx'} - D_{xy} -  D_{yx'} + 1\right\rangle \,,\nonumber\\
    &= C_F\Xi_{\rm LO}(\xt,\yt,\xt')+\mathcal{O}(Q_s^2r_{zy}^2)
\end{align}
After all these approximations and going to momentum space, the cross-section (plus its complex conjugate) can be cast into the simpler form
\begin{align}
 \left.\frac{\der \sigma^{\gamma_{\rm T}^{\star}+A\to q+X}}{ \der^2\bt\der^2 \Pt}\right|_{\overline{\rm R1}} &=\frac{\alpha_{\rm em}e_f^2N_c}{(2\pi)}\frac{\alpha_s C_F}{\pi^2} \int\der^2\qt \mathcal{D}(x,\qt) \int_0^1\der z\int_{x_\star  P_\perp^2/Q^2}^1\frac{\der z_g}{z_g}\left[1+(1-z_g)^2\right]\nonumber\\
 &\times\int\frac{\der^2\ltthre}{2\pi}\left[\mathcal{H}_{\overline{\rm R1}}^{ij*} \mathcal{H}_{\overline{\rm R1}}^{ij}-\mathcal{H}_{\overline{\rm R1},\rm uv}^{ij*} \mathcal{H}_{\overline{\rm R1},\rm uv}^{ij}\right]
\end{align}
with the hard factor of diagram $\overline{\rm R1}$
\begin{align}
\mathcal{H}_{\overline{\rm R1}}^{ij}&\equiv \int\frac{\der^2\rzyt}{(2\pi)}\int\frac{\der^2\rxyt}{(2\pi)}e^{-i\Pt\cdot\rxyt}\left(e^{i\qt\cdot\rxyt}-1\right)e^{i\ltthre\cdot\rzyt}\frac{\rzyt^i\rxyt^j}{\rzyt^2}\frac{\bar Q K_1\left(\bar Q\sqrt{\rxyt^2+\omega\rzyt^2}\right)}{\sqrt{\rxyt^2+\omega\rzyt^2}}\\
&=-\frac{\ltthre^i(\Pt^j-\qt)^j}{\left[(\Pt-\qt)^2+\bar Q^2\right]\left[\ltthre^2+\omega((\Pt-\qt)^2+\bar Q^2)\right]}+\frac{\ltthre^i\Pt^j}{\left[ P_\perp^2+\bar Q^2\right]\left[\ltthre^2+\omega( P_\perp^2+\bar Q^2)\right]}
\end{align}
and its UV regulator,
\begin{align}
    \mathcal{H}_{\overline{\rm R1},\rm uv}^{ij}&\equiv \int\frac{\der^2\rzyt}{(2\pi)}\int\frac{\der^2\rxyt}{(2\pi)}e^{-i\Pt\cdot\rxyt}\left(e^{i\qt\cdot\rxyt}-1\right)e^{i\ltthre\cdot\rzyt}\frac{\rzyt^i\rxyt^j}{\rzyt^2}\frac{\bar Q K_1\left(\bar Qr_{xy}\right)\exp\left(-\frac{\rzyt^2}{4\xi}\right)}{r_{xy}}\\
    &=\left[\frac{(\Pt-\qt)^j}{(\Pt-\qt)^2+\bar Q^2}-\frac{\Pt^j}{ P_\perp^2+\bar Q^2}\right]\frac{\ltthre^i}{\ltthre^2}\left(1-e^{-\xi\ltthre^2}\right)
\end{align}
In principle, one choose the arbitrary parameter $\xi$ in the UV regulator to be $\xi=\rxxtp^2e^{\gamma_E}/2$ to match the expression given by Eq.\,\eqref{eq:R1bar}. This makes the calculation more difficult, although in the dilute limit, one is allowed to use $\xi = c_0/ P_\perp^2$ instead (it gives identical finite pieces up to powers of $Q_s^2/ P_\perp^2$ corrections).
Integrating over the auxiliary variable $\ltthre$ gives
\begin{align}
    &\int\frac{\der^2\ltthre}{2\pi}\left[\mathcal{H}_{\overline{\rm R1}}^{ij*} \mathcal{H}_{\overline{\rm R1}}^{ij}-\mathcal{H}_{\overline{\rm R1},uv}^{ij*} \mathcal{H}_{\overline{\rm R1},uv}^{ij}\right]=\frac{(\Pt-\qt)^2}{2\left[(\Pt-\qt)^2+\bar Q^2\right]^2}\left[-1+\ln(c_0)-\ln\left(\xi(\omega((\Pt-\qt)^2+\bar Q^2))\right)\right]\nonumber\\
    &+\frac{ P_\perp^2}{2\left[ P_\perp^2+\bar Q^2\right]^2}\left[-1+\ln(c_0)-\ln\left(\xi(\omega( P_\perp^2+\bar Q^2))\right)\right]-\frac{2\Pt\cdot(\Pt-\qt)}{2\left[(\Pt-\qt)^2+\bar Q^2\right]\left[ P_\perp^2+\bar Q^2\right]}\Bigg[\ln(c_0)-\ln(\omega\xi)\nonumber\\
    &-\frac{\left[(\Pt-\qt)^2+\bar Q^2\right]\ln((\Pt-\qt)^2+\bar Q^2)-\left[ P_\perp^2+\bar Q^2\right]\ln( P_\perp^2+\bar Q^2)}{(\Pt-\qt)^2- P_\perp^2}\Bigg]
\end{align}
Finally, the take the dilute limite $ P_\perp^2\gg q_\perp^2$ and we perform the integral over $z_g$ and $z$ (which is dominated by the endpoint $z=0$ so that the upper limit of the $z$ integration is irrelevant):
\begin{align}
     \left.\frac{\der \sigma^{\gamma_{\rm T}^{\star}+A\to q+X}}{  \der^2 \Pt}\right|_{\overline{\rm R1}} =\left.\frac{\der \sigma^{\gamma_{\rm T}^{\star}+A\to q+X}}{  \der^2 \Pt}\right|_{\rm LO}\times\frac{\alpha_s C_F}{\pi}&\left[-\frac{3}{4}+\frac{\pi^2}{12}+\frac{3}{8}\ln\left(\frac{Q^2}{ P_\perp^2}\right)-\frac{1}2\ln\left(\frac{Q^2}{x_\star P_\perp^2}\right)\ln\left(\frac{Q^2}{ P_\perp^2}\right)+\frac{1}{4}\ln^2\left(\frac{Q^2}{x_*  P_\perp^2}\right)\right.\nonumber\\
     &\left.-\frac{5}{6}\left(-\frac{3}{4}+\ln\left(\frac{Q^2}{x_\star  P_\perp^2}\right)\right)\right]\label{eq:R1b-final}
\end{align}
Like the UV regularised diagram $\rm SE1\times LO^*$, the UV regularised diagram $\overline{\rm R1}\times \overline{\rm R1}^*$ generates a Sudakov double logarithm with coefficient $-\alpha_s C_F/(4\pi)\ln^2(Q^2/ P_\perp^2)$. This expression also illustrates our claim in the beginning of the section dedicated to the real diagrams: although the amplitude $\overline{\rm R1}$ is a real amplitude, the transverse momentum of the quark jet in Eq.\,\eqref{eq:R1b-final} comes from the initial state quark TMD (as encoded in the LO cross-section), so it effectively plays the role of \textit{virtual} correction to the evolution equation of the quark TMD discussed in the main text of the Letter.
The comment that we made earlier after \eqn{eq:SE1-final}, concerning the cancellation of the mixed double logs of the type \eqref{eq:mix}, also applies to the above equation.

\subsubsection{2.4.2 Real gluon emission from the measured quark before the shock-wave}

Now, we discuss the real gluon emission by the measured quark  before its scattering with  the SW, denoted 
as $\rm R1\times \rm R1^*$. Its   expression can be found in \cite{Caucal:2024cdq}:
\begin{align}
    \left.\frac{\der \sigma_{\rm CGC}^{\gamma_{\rm T}^{\star}+A\to q+X}}{  \der^2\bt\der^2 \Pt \der z}\right|_{\rm R1\times R1^*} &=\frac{\alpha_{\rm em}e_f^2N_c}{(2\pi)^4}\int\der^2\rxyt\der^2\rxpyt\der^2\rzyt e^{-i\Pt \cdot \rxxtp} \frac{\alpha_s}{2\pi^2} Q^2   \Xi_{\rm NLO,4}(\xt,\yt,\zt,\xt')  \nonumber\\
    & \times \int_0^{1-z} \frac{\der z_g}{z_g}\left\{\frac{K_1(QX_R)K_1(QX_R')}{X_R X_R'}\left[\bar z^2(z^2+(1-\bar z)^2)\left(\bar z^2 + (1 - \bar z)^2\right)\frac{\rzxt\cdot\rzxpt}{\rzxt^2\rzxpt^2}\RtR\cdot\RtR'\right.\right.\nonumber\\
    &\left.\left.+\bar z^2z_g(1+z-\bar z)(1-2\bar z)\frac{\rzxt\times\rzxpt}{\rzxt^2\rzxpt^2}\RtR\times\RtR'\right]\right\}_{z_\star}
\end{align}
with $\RtR=\rxyt+z_g/(z+z_g)\rzxt$.
Using similar techniques, the leading power piece of this diagram at the cross-section level is given by
\begin{align}
 \left.\frac{\der \sigma^{\gamma_{\rm T}^{\star}+A\to q+X}}{  \der^2\bt\der^2 \Pt}\right|_{\rm R1} &=\frac{\alpha_{\rm em}e_f^2N_c}{2\pi} \frac{\alpha_s C_F}{\pi^2} \int\der^2\qt \mathcal{D}(x,\qt) \int_0^1\der z\int_{x_\star P_\perp^2/Q^2}^1\frac{\der z_g}{z_g}\left[z_g^2+(1-z_g)^2\right]\int\frac{\der^2\ltthre}{(2\pi)}\mathcal{H}_{\rm R1}^{ij*} \mathcal{H}_{\rm R1}^{ij}
\end{align}
Noting $\Delta=\sqrt{\omega}\bar Q$,
\begin{align}
\mathcal{H}_{\rm R1}^{ij}&\equiv \int\frac{\der^2\rzyt}{(2\pi)}\int\frac{\der^2\rxyt}{(2\pi)}e^{-i\Pt\cdot\rxyt}\left(e^{i\qt\cdot\rxyt}-1\right)e^{i\ltthre\cdot\rzyt}\frac{\rzyt^i\rxyt^j}{\rxyt^2}\frac{\Delta K_1\left(\Delta \sqrt{\rzyt^2+\rxyt^2/\omega}\right)}{\sqrt{\rzyt^2+\rxyt^2/\omega}}\\
&=-\frac{\ltthre^i(\Pt^j-\qt)^j}{\left[\ltthre^2/\omega+\bar Q^2\right]\left[\ltthre^2+\omega((\Pt-\qt)^2+\bar Q^2)\right]}+\frac{\ltthre^i\Pt^j}{\left[\ltthre^2/\omega+\bar Q^2\right]\left[\ltthre^2+\omega( P_\perp^2+\bar Q^2)\right]}
\end{align}
The integral over $\ltthre$ is straightforward, and one gets
\begin{align}
    \int\frac{\der^2\ltthre}{(2\pi)}\mathcal{H}_{\rm R1}^{ij*} \mathcal{H}_{\rm R1}^{ij}&=\frac{-2(\Pt-\qt)^2+((\Pt-\qt)^2+2\bar Q^2)\ln\left(1+\frac{(\Pt-\qt)^2}{\bar Q^2}\right)}{2(\Pt-\qt)^4}+\frac{-2 P_\perp^2+( P_\perp^2+2\bar Q^2)\ln\left(1+\frac{ P_\perp^2}{\bar Q^2}\right)}{2\Pt^4}\nonumber\\
    &-\frac{(\Pt-\qt)\cdot\Pt}{(\Pt-\qt)^4\Pt^4((\Pt-\qt)^2- P_\perp^2)}\Bigg[(\Pt-\qt)^2\Pt^4- P_\perp^2(\Pt-\qt)^4\nonumber\\
    &-\Pt^4((\Pt-\qt)^2+\bar Q^2)\ln\left(1+\frac{(\Pt-\qt)^2}{\bar Q^2}\right)+(\Pt-\qt)^4( P_\perp^2+\bar Q^2)\ln\left(1+\frac{ P_\perp^2}{\bar Q^2}\right)\Bigg]
\end{align}
This diagram does not generate Sudakov double logarithm, but it contributes to the Sudakov single logarithm. Expanding for $P_\perp \gg q_\perp$, we get 
\begin{align}
    \left.\frac{\der \sigma^{\gamma_{\rm T}^{\star}+A\to q+X}}{  \der^2 \Pt}\right|_{\rm R1}&= \left.\frac{\der \sigma^{\gamma_{\rm T}^{\star}+A\to q+X}}{  \der^2 \Pt}\right|_{\rm LO}\times \frac{(-\alpha_s)C_F}{8\pi}\left[1-\ln\left(\frac{Q^2}{x_\star P_\perp^2}\right)\right]\label{eq:R1-final}
\end{align}

\subsubsection{2.4.3 Interferences between real gluon emission before the shock-wave from the measured quark and the unmeasured antiquark}

Let's finally discuss the interferences between emissions by the quark and the antiquark, such as $\rm R1\times \overline{R1}^*$. The interference contribution reads
\begin{align}
    \left.\frac{\der \sigma_{\rm CGC}^{\gamma_{\rm T}^{\star}+A\to q+X}}{  \der^2\bt\der^2 \Pt \der z}\right|_{\rm R1\times \overline{R1}^*} &=\frac{\alpha_{\rm em}e_f^2N_c}{(2\pi)^4}\int\der^2\rxyt\der^2\rxpyt\der^2\rzyt e^{-i\Pt \cdot \rxxtp} \frac{\alpha_s}{2\pi^2}2zQ^2  \Xi_{\rm NLO,4}(\xt,\yt,\zt,\xt') \nonumber\\
    & \times   \int_0^{1-z} \frac{\der z_g}{z_g}\left\{\frac{K_1(QX_R)K_1(QX_R')}{X_R X_R'}  \left[\bar z(z(1-z)+\bar z(1-\bar z))(z(1-\bar z) + \bar z (1-z))\frac{\rzxt\cdot\rzyt}{\rzxt^2\rzyt^2}\RtR\cdot\RtRb'\right.\right.\nonumber\\
    &\left.\left.-\bar zz_g(z-\bar z)^2\frac{\rzxt\times\rzyt}{\rzxt^2\rzyt^2}\RtR\times\RtRb'\right]  \right\}_{z_\star}
\end{align}
where we recall that $\bar z=1-z-z_g$.
In the limit $r_{zy}\ll r_{xy}$ and $z\ll 1$, the two terms nicely combine to give
\begin{align}
    2z^2(1-z_g)^3z_g\frac{(\rxyt\cdot\rzyt)(\rzyt\cdot\rxpyt)-(\rxyt\times\rzyt)(\rzyt\times\rxpyt)}{\rxyt^2\rzyt^2}&=2z^2(1-z_g)^3z_g\frac{\rxyt\cdot\rxpyt}{\rxyt^2}
\end{align}
Including the complex conjugate diagram and using the same approximations as above, we find
\begin{align}
 \left.\frac{\der \sigma^{\gamma_{\rm T}^{\star}+A\to q+X}}{ \der^2\bt \der^2 \Pt}\right|_{\rm int} &=\frac{\alpha_{\rm em}e_f^2N_c}{2\pi} \frac{2\alpha_s C_F}{\pi^2} \int\der^2\qt\mathcal{D}(x,\qt) \int_0^1\frac{\der z}{z}\int_{x_\star  P_\perp^2/Q^2}^1\frac{\der z_g}{z_g}(1-z_g)^3z_g\int\frac{\der^2\ltthre}{(2\pi)}\mathcal{H}_{\rm int,a}^i\mathcal{H}_{\rm int,b}^{i*}
\end{align}
with the hard factor defined as
\begin{align}
\mathcal{H}^i_{\rm int,a}&\equiv \int\frac{\der^2\rzyt}{(2\pi)}\int\frac{\der^2\rxyt}{(2\pi)}e^{-i\Pt\cdot\rxyt}\left(e^{i\qt\cdot\rxyt}-1\right)e^{i\ltthre\cdot\rzyt}\frac{\rxyt^i}{\rxyt^2}\frac{\bar Q K_1\left(\bar Q \sqrt{\rxyt^2+\omega\rzyt^2}\right)}{\sqrt{\rxyt^2+\omega\rxyt^2}}\\
&=\frac{(-i)}{2\omega}\left[\frac{\Pt^i-\qt^i}{(\Pt-\qt)^2}\ln\left(1+\frac{(\Pt-\qt)^2}{\bar Q^2+\ltthre^2/\omega}\right)-\frac{\Pt^i}{ P_\perp^2}\ln\left(1+\frac{ P_\perp^2}{\bar Q^2+\ltthre^2/\omega}\right)\right]\\
\mathcal{H}^i_{\rm int,b}&\equiv \int\frac{\der^2\rzyt}{(2\pi)}\int\frac{\der^2\rxyt}{(2\pi)}e^{-i\Pt\cdot\rxyt}\left(e^{i\qt\cdot\rxyt}-1\right)e^{i\ltthre\cdot\rzyt}\rxyt^i\frac{\bar Q K_1\left(\bar Q \sqrt{\rxyt^2+\omega\rzyt^2}\right)}{\sqrt{\rxyt^2+\omega\rxyt^2}}\\
&=\frac{-2i\omega(\Pt-\qt)^i}{\left[\ltthre^2+\omega((\Pt-\qt)^2+\bar Q^2\right]^2}+\frac{2i\omega\Pt^i}{\left[\ltthre^2+\omega( P_\perp^2+\bar Q^2\right]^2}
\end{align}
The $\ltthre$ integration is again straightforward since
\begin{align}
 &\int\frac{\der^2\ltthre}{(2\pi)}\mathcal{H}_{\rm int,a}^i\mathcal{H}_{\rm int,b}^{i*}= \frac{ P_\perp^2-\bar Q^2\ln\left(1+\frac{ P_\perp^2}{\bar Q^2}\right)}{2\omega P_\perp^2( P_\perp^2+\bar Q^2)}+\frac{(\Pt-\qt)^2-\bar Q^2\ln\left(1+\frac{(\Pt-\qt)^2}{\bar Q^2}\right)}{2\omega(\Pt-\qt)^2((\Pt-\qt)^2+\bar Q^2)}\nonumber\\
 &+\frac{\Pt\cdot(\Pt-\qt)\bar Q^2}{2\omega(\Pt-\qt)^2 P_\perp^2\left[ P_\perp^2+\bar Q^2\right]}\ln\left(1+\frac{(\Pt-\qt)^2}{\bar Q^2}\right)+\frac{\Pt\cdot(\Pt-\qt)}{2\omega P_\perp^2\left[ P_\perp^2-(\Pt-\qt)^2)\right]}\ln\left(\frac{(\Pt-\qt)^2+\bar Q^2}{ P_\perp^2+\bar Q^2}\right)\nonumber\\
 &+\frac{\Pt\cdot(\Pt-\qt)\bar Q^2}{2\omega(\Pt-\qt)^2 P_\perp^2\left[(\Pt-\qt)^2+\bar Q^2\right]}\ln\left(1+\frac{ P_\perp^2}{\bar Q^2}\right)+\frac{\Pt\cdot(\Pt-\qt)}{2\omega(\Pt-\qt)^2\left[ P_\perp^2-(\Pt-\qt)^2)\right]}\ln\left(\frac{(\Pt-\qt)^2+\bar Q^2}{ P_\perp^2+\bar Q^2}\right)
\end{align}
Note that the additional $1/z$ factor in the $z_g$ integration combines with the $1/\omega=z/(z_g(1-z_g)$ to give a leading power result. In particular, in the dilute limit,
\begin{align}
    \left.\frac{\der \sigma^{\gamma_{\rm T}^{\star}+A\to q+X}}{  \der^2 \Pt}\right|_{\rm int}&= \left.\frac{\der \sigma^{\gamma_{\rm T}^{\star}+A\to q+X}}{  \der^2 \Pt}\right|_{\rm LO}\times \frac{\alpha_sC_F}{4\pi}\left[-\frac{3}{2}+\ln\left(\frac{Q^2}{x_\star P_\perp^2}\right)\right]\label{eq:int-final}
\end{align}

\subsection{2.5 Summary}

To sum up this lengthy calculation, we combine together all our intermediate results for the NLO diagrams which contribute at leading twist, namely Eq.\,\eqref{eq:Adef-final}-\eqref{eq:Bdef-final}, Eq.\,\eqref{eq:SE1-final}, Eq.\,\eqref{eq:V1-final}, Eq.\,\eqref{eq:R1b-final}, Eq.\,\eqref{eq:R1-final} and Eq.\,\eqref{eq:int-final}. For the clustering condition A, we get
\begin{align}
      \left.\frac{\der\sigma^{\gamma^*_{\rm T}+A\to j+X}}{\der^2\Pt}\right|_{\rm NLO, A-def}=\left.\frac{\der\sigma^{\gamma^*_{\rm T}+A\to j+X}}{\der^2\Pt}\right|_{\rm LO}& \times  \frac{\alpha_sC_F}{\pi}\left[-\frac{3}{4}\ln^2\left(\frac{Q^2}{P_\perp^2}\right)+\left(\frac{3}{4}-\ln(R)\right)\ln\left(\frac{Q^2}{P_\perp^2}\right)\right.\nonumber\\
      &\left.-\frac{3}{2}\ln(2R)+\frac{17}{4}-\frac{5\pi^2}{12}+\frac{3}{4}\ln^2(x_\star)+\frac{3}{8}\ln(x_\star)+\mathcal{O}(R^2)\right]\,.\label{eq:virtual-xs-Adef}
\end{align}
while for the clustering condition B,
\begin{align}
      \left.\frac{\der\sigma^{\gamma^*_{\rm T}+A\to j+X}}{\der^2\Pt}\right|_{\rm NLO, B-def}=\left.\frac{\der\sigma^{\gamma^*_{\rm T}+A\to j+X}}{\der^2\Pt}\right|_{\rm LO}& \times  \frac{\alpha_sC_F}{\pi}\left[-\frac{1}{4}\ln^2\left(\frac{Q^2}{P_\perp^2}\right)+\left(\frac{3(1-\beta/2)}{4}+\ln(R)\right)\ln\left(\frac{Q^2}{P_\perp^2}\right)\right.\nonumber\\
  &\left.-\frac{3}{2}\ln(R)+\frac{11}{4}-\frac{3\pi^2}{4}+\frac{3}{4}\ln^2(x_\star)+\frac{3}{8}\ln(x_\star)+\mathcal{O}(R^2)\right]\,.\label{eq:virtual-xs-Bdef}
\end{align}
An important consistency check of these results is the fact that they do not involve mixed double logarithms of the type shown in \eqn{eq:mix}. This is important since it guarantees that the small-$x$ and the Sudakov resummations (equivalently, the B-JIMWLK and the CSS evolutions) can be performed simultaneously without overlap in phase space. In particular, this property is consistent with the fact that the CSS evolution is local in $x$. Moreover, the $x_\star$--dependent remainder is identical for both jet definitions; this shows that small-$x$ evolution is not sensitive to the final state collinear dynamic. In both cases, we recognise the same term of the form $-(3/2)\ln(R)$ coming from final-state, hard and collinear, gluon splittings.

Let us now turn to the differences between the two classes of jet definitions, $A$ and $B$, as 
manifest from the inspection of the Sudakov logarithms in Eqs.~\eqref{eq:virtual-xs-Adef} and \eqref{eq:virtual-xs-Bdef}.  Clearly, the main difference
refers to the coefficient of the  Sudakov double logarithm. With the clustering condition $B$, this coefficient is equal to  $-\alpha_sC_F/(4\pi)$ 
(like in Eq.~(12) in the Letter) and is consistent with TMD factorisation, as also explained in the Letter: \texttt{(i)} it has the right value to cancel the (single) real Sudakov log in Eq.~(11) of the Letter after integrating over $P_\perp$, and \texttt{(ii)} it is furthermore consistent with expectations from the CSS formalism~\cite{Collins:1981uk,Collins:1981uw,Collins:1984kg,Collins:2011zzd}.  On the contrary, the respective coefficient predicted by the clustering condition $A$, that is, $-3\alpha_sC_F/(4\pi)$, not only is inconsistent with the general CSS results aforementioned, but also does not fulfil the condition of real--virtual cancellation after integrating over $P_\perp$; indeed, with this jet definition, the real Sudakov logarithm is found to be twice as large as that shown in the second line of Eq.~(11), hence it cannot cancel against a virtual double logarithm which is {\it three} times larger than that in \eqn{eq:virtual-xs-Bdef}.

As furthermore explained in the Letter, the coefficient of the Sudakov {\it single} logarithm is important too to ensure TMD factorisation, namely it needs to match the quark anomalous dimension $\Gamma_q=\frac{3\alpha_sC_F}{4\pi}$ and by the same token to provide the correct virtual piece of the splitting function  $\mcal{P}_{qq}(\xi)$ (recall the discussion after Eq.~(13) in the Letter). By inspection  of  \eqn{eq:virtual-xs-Bdef}, it is clear that these conditions are satisfied if and only we choose $\beta=0$. We recall that in the narrow jet approximation ($R\ll 1$), $\beta=1$ corresponds to the inclusive $k_t$ algorithm designed in~\cite{Catani:1992zp}, while $\beta=2$ corresponds the inclusive generalised-$k_t$ algorithm for $e^+e^-$ implemented in the Breit frame. Yet, in both cases, these widely used jet definitions fail to correctly account for the single Sudakov logarithm.
In particular,  with $\beta=2$, the respective coefficient simply vanishes!

From a physical point of view, the case $\beta=2$ can be understood as follows. Consider the situation where, in the dipole frame, the tagged quark is slow, with $z_1\ll 1$. That is, the fermion struck by the photon in the Breit frame is the (unmeasured) antiquark. Still in the Breit frame, the tagged jet  propagates in the fragmentation region of the nuclear target. If this jet is reconstructed  using the inclusive generalised-$k_t$ algorithm for $e^+e^-$ (i.e.
\eqn{eq:cluster-DIS} with $\beta=2$) and if the gluon is close in angle to the tagged quark with $z_1\ll 1$, then both particles are recombined into the same jet with $z=z_g+z_1\ll 1$. In such a case,  the gluon emission does not modify the structure of the final state, so there is a complete cancellation of the single Sudakov logarithms between real (in-cone) and virtual emissions, cf. \eqn{eq:virtual-xs-Bdef} with $\beta=2$. On the other hand,
when using the distance measure given by Eq.\,(5) in the Letter with $p=1$   (or \eqn{eq:cluster-DIS} with $\beta=0$), the gluon and the quark jet with $z_g,z_1\ll 1$ are strongly penalised by the clustering criterion: the left hand side of \eqn{eq:cluster-DIS} becomes large for $z_1 z_g \ll 1$, while its r.h.s. is equal to 1 for $\beta=0$. So, these very soft quark-gluon pairs are typically not  recombined inside the same jet by our new clustering condition, meaning that the virtual contributions to the single Sudakov logs remain uncompensated, as indeed visible in  \eqn{eq:virtual-xs-Bdef} when $\beta=0$.

To summarise, the physical jet definition for jets in SIDIS --- the one to be consistent with TMD factorisation at NLO --- is the one that we propose 
 in Eq.~(5) in the Letter with the choice $p=1$, or, equivalently, that shown in \eqn{eq:cluster-DIS} of this Supplemental Material, with the choice $\beta=0$.
 For this definition, the result that we have just derived for the NLO virtual corrections, i.e.  \eqn{eq:virtual-xs-Bdef}  with $\beta=0$, coincides indeed with that shown in Eq.~(12) in the Letter.

 After adding together the virtual corrections as given by \eqn{eq:virtual-xs-Bdef}  with $\beta=0$ (equivalently, Eq.~(13) in the Letter) and the real corrections shown in Eq.~(11) from the Letter, we are finally in a position to exhibit the NLO result for the jet measurement in SIDIS in the approximations of interest. This is shown in Eq.~(15) from the Letter, that we repeat here for convenience:
 \begin{align}
    \frac{\der\sigma^{\gamma^*_{\rm T}+A\to j+X}}{\der^2\Pt}\Big|_{\rm NLO}&=\frac{8\pi^2\alpha_{\rm em}e_f^2}{Q^2} x\mathcal{F}_{q}(x, \Pt,Q^2)
    \left[1-\frac{3\alpha_sC_F}{2\pi}\ln(R)+\mathcal{O}(\alpha_s)\right]\,,\label{eq:NLO-SIDIS-final}
\end{align}
where it is understood that
\begin{align}
 x\mathcal{F}_{q}(x, \Pt,Q^2)= x\mathcal{F}_{q}^{(0)}(x, \Pt)+x\mathcal{F}_{q}^{(1)}(x,\Pt, Q^2)\Big|_{R}+x\mathcal{F}_{q}^{(1)}(x,\Pt, Q^2)\Big|_{V}\,.
\end{align}
The (real and virtual) NLO corrections exhibit logarithmic dependences upon the hard scale $Q^2$, as visible in Eqs.~(11) and (13). By taking a derivative w.r.t. $\ln Q^2$, one can promote these one-loop corrections to an evolution equation, which reads
\begin{align}
	\label{CSS}
 \frac{\del\, x \mcal{F}_q(x, \Pt, Q^2)}{\del \ln Q^2}=
& \, \frac{C_F}{2\pi}\left\{\frac{\alpha_s(\PT^2) }{\PT^2}
\int^{P_\perp^2}_{\Lambda^2}\rmd \ell_\perp^2\,x\mcal{F}_q(x, \bm{\ell}_{\perp}, Q^2)
-\int_{P_\perp^2}^{Q^2}\frac{\rmd \ell_\perp^2}{\ell_\perp^2}\alpha_s(\ell_\perp^2)\,x\mcal{F}_q(x, \Pt, Q^2)\right\}\nonumber\\
&+\frac{3}{2}\frac{\alpha_s(Q_\perp^2)C_F}{\pi}\,x\mcal{F}_q(x,\Pt, Q^2)\,. \end{align}
In all the terms, $\alpha_s$ runs with the transverse momentum of the daughter gluons produced by the hard splitting. 

As mentioned in the Letter, \eqn{CSS} can be recognised as the diagonal version of the Collins-Soper-Sterman (CSS) equations~\cite{Collins:1981uk,Collins:1981uw,Collins:1984kg,Collins:2011zzd}. (The same equations have also been obtained with the SCET formalism~\cite{Becher:2010tm,Becher:2012yn,Echevarria:2011epo,Echevarria:2014rua,Chiu:2012ir,Ebert:2019tvc}.) That is, this is the equation obeyed by the quark TMD $x\mathcal{F}_q^{\rm (sub)}(x,\Pt,\mu_F,\zeta_c)$ provided the two resolution scales are identified with each other and with the hard scale of our problem, which is $Q$: $\mu_F=\zeta_c=Q$.

\eqn{CSS} should be solved as an initial value problem with the initial condition at $Q^2=P_\perp^2$ provided by Eq.~(11) in the Letter, that is 
\begin{align}
 x\mathcal{F}_{q}(x, \Pt,Q^2=P_\perp^2)&\, = x\mathcal{F}_{q}^{(0)}(x, \Pt)+
 \frac{\alpha_s}{2\pi^2}\frac{1}{P_\perp^2}\!\int_{x}^{1}\! {\der \xi} 
     P_{qq}^{(+)}(\xi) \frac{x}{\xi}f_q\!\left(\frac{x}{\xi},P_\perp^2\right)\nonumber\\
&\,=\,\frac{1}{\pi} \frac{\del xf_q(x, P_\perp^2)}{\del P_\perp^2}\,-\,\frac{3}{2}
 \frac{\alpha_s C_F}{2\pi^2}\,\frac{xf_q(x, P_\perp^2)}{P_\perp^2}
\,, \end{align}
where the second line follows after using the DGLAP equation, as shown in Eq.~(14) in the Letter.
Hence, in practice, one should first solve the DGLAP equation for the quark PDF $xf_q(x, P_\perp^2)$ and then use its solution to build the initial condition for the CSS equation Eq.~\eqref{CSS}.

The impact-parameter version of the CSS equation is obtained after a Fourier transform from $\Pt$ to $\bt$, 
\beq\label{FTTMD}
 x\tilde{\mathcal{F}}_{q}(x,  \bt,  Q^2)\,\equiv\int \frac{\rmd^2\Pt}{(2\pi)^2}\,\rme^{i\Pt\cdot\bt}\, x\mcal{F}_q(x, \Pt, Q^2)\,,
\eeq
and takes the expected from (the diagonal version of the CSS equations in $b_\perp$-space, see e.g. Eq.~(4.12) in \cite{Boussarie:2023izj})
\begin{align}\label{CSSimpact}
 \frac{\del x\tilde{\mathcal{F}}_{q}(x,  \bt,  Q^2)}{\del\ln Q^2}=&\frac{C_F}{\pi}
 \left[-\frac{1}{2}\int_{\mu_b^2}^{Q^2} \frac{\rmd \ell_\perp^2}{\ell_\perp^2}\, \alpha_s(\ell_\perp^2)
+\frac{3}{2}\alpha_s(Q^2) \right] x\tilde{\mathcal{F}}_{q}(x,  \bt,  Q^2),\end{align}
where we have introduced the notation $\mu_b^2\equiv c^2_0/\bt^2$  with  $c_0=2\,e^{-\gamma_E}\simeq 1.123$ for the lower limit. Notice that the real term from  the $P_\perp$-space equation Eq.~\eqref{CSS} has (formally) disappeared after going to $b_\perp$-space: the Fourier transform has ``corse-grained'' the modes with $P_\perp \lesssim \mu_b$, for which real and virtual corrections have mutually cancelled. So, we are only left with the virtual corrections at $P_\perp > \mu_b$. This explains the lower limit $\mu_b^2$ in the above integral over  $\ell_\perp^2$.

\eqn{CSSimpact} can be easily solved for an arbitrary initial condition $x\tilde{\mcal{F}}_0(x, \bt^2)\equiv 
x\tilde{\mathcal{F}}_{q}(x,  \bt,  Q^2=\mu_b^2)$. However, the inverse Fourier transform to momentum-space is sensitive to very large values of $b_\perp$, for which a non-perturbative prescription is generally needed. Such prescriptions are well documented in the literature \cite{Collins:2011zzd,Boussarie:2023izj}. Here, we would only like to add that the sensitivity to  non-perturbative physics is alleviated by the phenomenon of gluon saturation: for sufficiently small values of $x$ and/or large values of the atomic number $A$, the saturation momentum is a semi-hard scale, $Q_s^2(x, A)\gg \Lambda^2$, and the initial condition  $x\tilde{\mcal{F}}_0(x, \bt^2)$ taken from the CGC effective theory, cf.~\eqn{eq:quarkTMD-def}, is rapidly decreasing at large distances, $b_\perp\gg 1/Q_s$, thus ensuring the convergence of the inverse Fourier transform.

\section{3 General Discussion on TMD Factorisation for SIDIS with the Jet Algorithm from Eq.~(5)}

In this section, we shall present an alternative argument for TMD factorisation, which is formulated in the more traditional framework of the target picture --- that is, it follows the dynamics of the quark struck by the virtual photon. This argument neglects the saturation effects, hence it applies for  moderate values of $x$, and also at small $x$ but sufficiently large values for the jet  transverse momentum $P_\perp$, within the range $Q^2\gg P_\perp^2\gg Q_s^2(x)$. In this framework, one can rely on well established techniques to compute the NLO corrections to both the SIDIS cross-section and the quark TMD (starting with its operator definition). Previous calculations of SIDIS have focused on the measurement of a hadron (see for instance, the QCD factorisation and one-loop calculations in Refs.~\cite{Ji:2004wu,Collins:2011zzd,Sun:2013hua}). Here, we shall rather consider the production of a jet, with the purpose of demonstrating that --- in this context too --- the proper definition of the jet is essential in order to achieve TMD factorisation. Namely, we will show that the NLO corrections to the jet cross-section match the expected DGLAP+CSS evolution of the quark TMD if and only if one uses the jet definition given by Eq.\,(5) in the main text (or, equivalently, Eq.\,\eqref{eq:cluster-DIS} with $\beta=0$ in this Supplemental Material).

Consider  jet production in SIDIS, that is, the process $e+A\to e' + \textrm{jet} (P_\perp) + X$, in the (hard) TMD kinematics
at $Q^2\gg P_\perp^2\gg Q_s^2(x)$. We shall parametrise the corresponding cross-section as follows,
\begin{eqnarray}
    \frac{\der^4 \sigma^{e+A\to e'+j+X}}{\der x \der y\der^2\Pt}
      &=& \sigma_0^{\rm (DIS)} \, x F_{UU}(x,\Pt,Q)
     \ ,
\end{eqnarray}
where $x\equiv x_{\rm Bj}$,  $\Pt$ is the transverse momentum of the produced jet with respect to the virtual photon direction,
 and (with the usual DIS kinematic variables $y$, $x_{\rm Bj}$, $Q^2$, and $W^2$, and $S_{ep}=(\ell_e+P_N)^2$)
\beq
\sigma_0^{\rm (DIS)}=\,\frac{4\pi\alpha^2_{\rm em}e_f^2  
S_{ep}}{Q^4}\, (1-y+y^2/2)\,.
\eeq
 Note that with respect to Eq.\,(1) in the main text, we include here the leptonic part of the cross-section. 
  $F_{UU}$ represents the spin-averaged structure function and. At LO, it is given by the (LO) quark TMD, cf. 
  Eq.\,(1) in the Letter: $xF_{UU}^{(0)}=x\mathcal{F}^{(0)}_q(x,\Pt)$.
  
  As announced, we consider this process in the target picture and 
  assume that the LO approximation corresponds to a valence quark, which is collinear with the nuclear target, $k_q^\mu=x P_N^\mu$, so in particular it  has zero intrinsic transverse momentum:
  \beq\label{eq:FUU0}
  xF_{UU}^{(0)}(x,\Pt)\,=\,\delta^{(2)}(\Pt)\, xf_q^{(0))}(x),\eeq
  where $x f_q^{(0)}(x)$ represents the integrated quark distribution in the parton model and is scale independent.

  For the purposes of the TMD factorisation,  and notably in order to deal with  low values for the transverse momentum,
  it is more convenient to work in the impact parameter representation, as obtained after a 
 Fourier transform:  
\begin{eqnarray}
xF_{UU}(x,\Pt,Q)&=&\int \frac{\der^2\bt}{(2\pi)^{2}} e^{i \Pt\cdot \bt }x\widetilde
{F}_{UU}(x,\bt,Q)  \ .\label{eq:FT-FUU}
\end{eqnarray}
The \eqn{eq:FUU0} implies, at tree-level, $x\widetilde {F}_{UU}(x,\bt,Q)$ is independent upon both $\bt$ and $Q^2$:
\begin{equation}
    x\tilde{F}^{(0)}_{UU}(x,\bt,Q)=
    xf_q^{(0)}(x).\label{eq:LO-FUU}
\end{equation}

\begin{figure}
    \centering
    \includegraphics[width=0.6\linewidth]{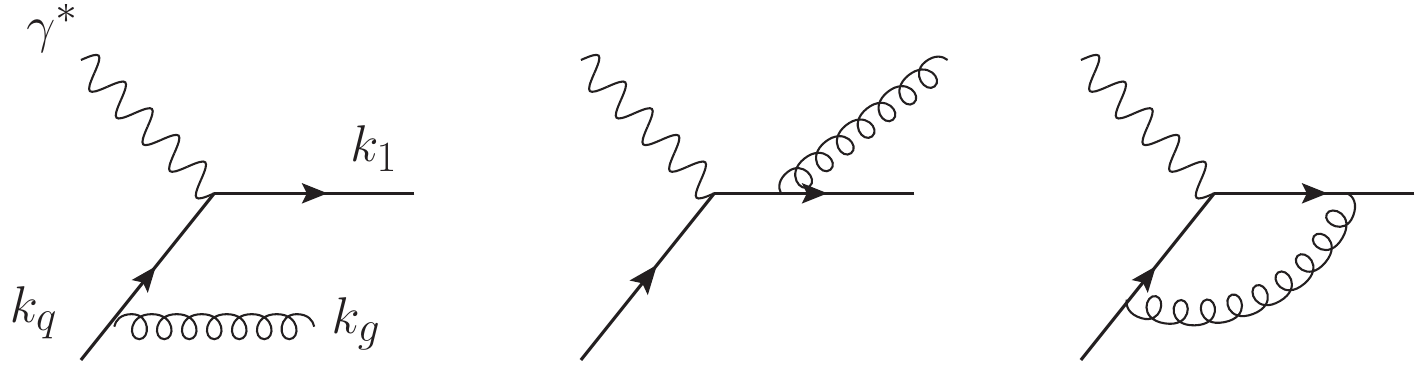}
    \caption{Feynman diagrams for the calculation of the NLO corrections to the SIDIS jet cross-section.}
    \label{fig:NLO-graphs-moderatex}
\end{figure}

Before we proceed, let us emphasise an interesting difference w.r.t. the previous calculation using the dipole picture: in the present case, the measured jet can only be generated by the struck quark, so it necessarily carries a large fraction $z_1\equiv (k_1\cdot P_N)/(q\cdot P_N)\simeq 1$ of the longitudinal momentum  of the virtual photon. This has two important implications: \texttt{(i)} in the final state, there is no jet in the fragmentation region of the nuclear target, and \texttt{(ii)} the jet cross-section has no sensitivity to the parameter $\beta$ which enters the clustering condition $B$, \eqn{eq:cluster-DIS}. Indeed, the jet has a longitudinal fraction $z\gtrsim z_1 \simeq 1$ and then the $\beta$--dependence of \eqn{eq:cluster-DIS} becomes irrelevant at NLO (altough it still matters for higher order calculations, starting at two loop order). This discussion also shows that, unlike in the previous scenario at small $x$ (where the measured jet can be sourced by either fermion from a sea quark-antiquark pair), the present calculation at moderate  $x$ cannot discriminate between different jet definitions from the class represented by \eqn{eq:cluster-DIS}. With this in mind, let us return to the calculation of the jet cross-section for the case where at tree-level we have just a valence quark.

Clearly, the LO (or tree-level) approximation corresponds to the familiar ``handbag'' diagram for the absorption of the virtual photon by a valence quark. The NLO corrections are associated with the three amplitude graphs depicted in Fig.~\ref{fig:NLO-graphs-moderatex}:
real gluon emission in the initial state, or in the final state (i.e. prior and respectively after the photon absorption) and the virtual correction to the quark-photon vertex. (As before, we work in dimensional regularisation, so the quark self-energies are identically zero.)
The virtual graph in Fig.~\ref{fig:NLO-graphs-moderatex} gives the following contribution in the $\overline{\rm MS}$ scheme:
\begin{equation}
{\left.  \tilde{F}_{UU}(x,\bt,Q)\right|_{\cal V}}=\frac{\alpha_s}{2\pi}C_F\left(\frac{\mu^2}{Q^2}\right)^\epsilon \left\{-\frac{2}{\epsilon^2}-\frac{3}{\epsilon}-8\right\} \, .\label{eq:FUU-virtual}
\end{equation}
This does not modify the transverse momentum dependence, leading to a constant contribution in $b_\perp$-space. 

Another contribution which brings no momentum dependence and hence can be effectively seen as a virtual correction, is the
emission of a real gluon in the final state (cf. the middle graph in Fig.~\ref{fig:NLO-graphs-moderatex}), but such the gluon and the outgoing quark are clustered inside the same jet. This is known as ``the jet contribution'' and is similar
to Eq.\,\eqref{eq:incone-def} in the small $x$ calculation of section 2. As detailed in the previous section, the calculation of this contribution is strongly sensitive to the jet definition. We compute this contribution by using the jet algorithm in Eq.~(5) of the 
main text and thus find
\begin{eqnarray}
J_q&=&\frac{\alpha_s}{2\pi}C_F\left(\frac{\mu^2}{Q^2}\right)^\epsilon\left\{
\frac{1}{\epsilon^2}+\frac{1}{\epsilon}\ln\frac{1}{R^2}+\frac{3}{2}\frac{1}{\epsilon}
+\frac{1}{2}\ln^2\frac{1}{R^2}
+\frac{3}{2}\ln\frac{1}{R^2}+\frac{13}{2}-\frac{2}{3}\pi^2\right\}.\label{eq:Jq-term}
\end{eqnarray}

Consider now the real gluon emissions where the gluon jet and the quark jet are well separated from each other (in the
sense of our jet definition, of course). This emissions provide the transverse momentum $\Pt$ of the final jet via their recoil.
For more clarity, we denote the 4-momenta of the incoming quark, the outgoing quark and the gluon as $k_q^\mu$, $k_1^\mu$,
and $k_g^\mu$, respectively. Clearly, $\bk_{q\perp}=0$ and $\bk_{1\perp}=\Pt=-\bk_{g\perp}$. To compute gluon radiation,
it is convenient to chose the physical polarisation of the radiated gluon along the direction of the incoming hadron:
$\varepsilon(k_g)\cdot k_q=0$. This is convenient since in this gauge there is no soft radiation by the incoming quark, but only by the outgoing one. If we chose the hadron to be a left mover (as throughout this paper), then $k_q$ has only a minus component,
$k_q^\mu=\delta^{\mu-} xP_N^-$, and 
the gauge condition becomes $A^+=0$ (the LC gauge of the virtual photon projectile). 

Consider first soft gluon emission by the final quark. In the eikonal approximation, the respective contribution to the cross-section factorises and is proportional to 
\begin{align}
g^2 C_F\,\frac{k_1^\mu}{2k_g\cdot k_1 -i\epsilon}\,\frac{k_1^\nu}{2k_g\dot k_1 +i\epsilon}\,\left(-g^{\mu \nu}+
\frac{k_g^\mu k_q^\nu+k_g^\nu k_q^\mu}{k_g\cdot k_q}\right)= g^2 C_F S_g(k_q,k_1,k_g), \label{eq:out}
\end{align}
with the notation
\begin{equation}
    S_g(k_q,k_1,k_g)\equiv \,\frac{2k_q\cdot k_1}{(k_q\cdot k_g) (k_1\cdot k_g)} \,\simeq\,\frac{4}{k_{g\perp}^2}\, ,
\end{equation}
where the final, approximate, equality holds since $k_q\cdot k_1=k_q^- k_1^+= x z_1P_N^- q^+$, 
$k_q\cdot k_g = xz_g P_N^- q^+$, and 
\beq 
2k_1\cdot k_g=\frac{z_g}{z_1}k_{1\perp}^2+\frac{z_1}{z_g}k_{g\perp}^2-2 \bk_{1\perp}\cdot \bk_{g\perp}\,=\,
\frac{z_1}{z_g}\left(\bk_{g\perp} -\frac{z_g}{z_1}\bk_{1\perp}\right)^2
\simeq\,\frac{z_1}{z_g}\,k_{g\perp}^2
\eeq
where we have used $z_g\ll z_1\simeq 1$ and $k_{1\perp}=k_{g\perp}=P_\perp$. The final result in \eqn{eq:out} is gauge
independent and can alternatively be interpreted in the covariant gauge as the ``antenna pattern'' produced via soft radiation from both the incoming and the outgoing quarks, including their interference.

To obtain the contribution of the final-state radiation to the jet cross-section,
we need to integrate out the phase space of $k_g$, while keeping its transverse momentum fixed to $-\Pt$ and excluding gluon radiation inside the jet:
\begin{align}
\mathcal{S}_{\rm out}&\equiv 4\pi\alpha_sC_F\int\frac{\der k_g^+\der^2\kgt}{2(2\pi)^3k_g^+}\, \delta^{(2)}(\kgt+\Pt)
\,S_g(k_q,k_1, k_g)\,\Theta \left ((k_1+k_g)^2>Q^2z_gR^2\right)\nonumber \\
&\simeq
 \frac{\alpha_s C_F}{\pi^2 }\frac{1}{P_\perp^2}   \int\frac{\der z_g}{z_g}\, \Theta \left (P_\perp^2
 >Q^2z_g^2 R^2\right)\ ,
\end{align}
where we have applied the jet algorithm of Eq.\,(5) with $p=1$.  Recalling the lower limit $z_g\gg P_\perp^2/Q^2$ on $z_g$
(the boundary of the phase-space for high-energy evolution), we obtain 
\begin{equation}
   \mathcal{S}_{\rm out}=\frac{\alpha_sC_F}{2\pi^2}\frac{1}{P_\perp^2} 
   \ln\left(\frac{Q^2}{R^2 P_\perp^2}\right).
\end{equation}
This is recognised as the real Sudakov logarithm previously obtained in the dipole picture, cf. the second line in Eq.~(11) of the Letter. This result is clearly sensitive to the value of the parameter $p$ in the jet definition,  Eq.\,(5) of the main text: for a different value of $p$, like $p=0$, the coefficient in front of the Sudakov logarithm would be different, which in turn would entail a mismatch between the real and the virtual NLO corrections, to be shortly computed.

To complete the calculation of the real NLO contribution to $F_{UU}$ in the approximation of interest, we need to add the
contribution of the collinear gluon radiation from the incoming quark. A standard calculation using dimensional regularisation yields
\begin{eqnarray}
xF_{UU}(x,\Pt,Q)\Big |_{\mcal{R}}&=&\frac{\alpha_s C_F}{2\pi^2}
\frac{1}{P_\perp^2}\int_{x}^1\der\xi\,\frac{x}{\xi}
f_q^{(0)}\left(\frac{x}{\xi}\right)\left\{\frac{1+\xi^2}{(1-\xi)_+}+\epsilon(1-\xi)+\delta(1-\xi)\left[\ln\frac{Q^2}{P_\perp^2}+\ln\frac{1}{R^2}\right]\right\}\, .\label{eq:Fuu-real}
\end{eqnarray}
As an important consistency check, we notice that this result coincides --- modulo the $\epsilon$ dependent term which vanishes in four dimension --- with the corresponding one obtained in the dipole picture, cf. Eq.\,(11) in the Letter.

As earlier mentioned, it is convenient to transform this result to impact parameter space, cf. \eqn{eq:LO-FUU}. This can be done with the help of the following identities:
\begin{align}
\mu^{2\epsilon}\int\frac{\der^{2-2\epsilon}\Pt}{(2\pi)^{2-2\epsilon}}e^{-i\bt\cdot\Pt}
    \frac{1}{P_\perp^2}&=\frac{\pi^{-1+\epsilon}}{4}\left(\frac{\bt^2\mu^2}{2}\right)^\epsilon\Gamma(-\epsilon)\\
    \mu^{2\epsilon}\int\frac{\der^{2-2\epsilon}\Pt}{(2\pi)^{2-2\epsilon}}e^{-i\bt\cdot\Pt}
    \frac{1}{P_\perp^2}\ln\left(\frac{Q^2}{P_\perp^2}\right)&=\frac{1}{(4\pi)^{1-\epsilon}}\left(\frac{\bt^2\mu^2}{4}\right)^\epsilon\left[\gamma_E+\ln\left(\frac{Q^2\bt^2}{4}\right)-\frac{\Gamma'(-\epsilon)}{\Gamma(-\epsilon)}\right]\Gamma(-\epsilon)
\end{align}

The full NLO correction to $x\widetilde{F}_{UU}(x,\bt,Q)$ can finally be obtained by combining 
 the Fourier transform of Eq.\,\eqref{eq:Fuu-real} with the virtual corrections in Eq.\,\eqref{eq:FUU-virtual} and Eq.\,\eqref{eq:Jq-term}.
 We thus find
\begin{align}
x\widetilde{F}_{UU}^{(1)}(x,\bt,Q)&=
\,\frac{\alpha_sC_F}{2\pi}\int_{x}^1\der\xi\,\frac{x}{\xi}f_q^{(0)}\left(\frac{x}{\xi}\right)\Bigg\{\left(-\frac{1}{\epsilon}+\ln\frac{\mu_b^2}{\mu^2}\right)\frac{1}{C_F}P_{qq}^{(+)}(\xi)+(1-\xi) +\nonumber\\
&+\delta(1-\xi)\left[-\frac{1}{2}\left(\ln\frac{Q^2}{\mu_b^2}\right)^2+
\left(\frac{3}{2}-\ln\frac{1}{R^2}\right)\ln\frac{Q^2}{\mu_b^2}
+\frac{1}{2}\ln^2\left(\frac{1}{R^2}\right)+\frac{3}{2}\ln\frac{1}{R^2}-\frac{3}{2}-\frac{2\pi^2}{3}\right]\Bigg\} \ ,\label{euu}
\end{align}
where $\mu_b=c_0/b_\perp$, $c_0=2e^{-\gamma_E}$, and $P_{qq}^{(+)}(\xi)=C_F\frac{1+\xi^2}{(1-\xi)_+}$ is the quark splitting kernel. The second term in the curly bracket of Eq.\,\eqref{euu}, just before the $\delta(1-\xi)$ piece, is a finite piece coming from the product of the $1/\epsilon$ pole from the Fourier transform of $1/q_\perp^2$ and the $\mathcal{O}(\epsilon)$ term in Eq.\,\eqref{eq:Fuu-real}.  Under the Fourier transform, it is legitimate to identify $\mu_b^2$ with $P_\perp^2$. With this identification, 
it is easy to see that the Sudakov double and single logarithms in the square brackets multiplying the $\delta$--function $\delta(1-\xi)$ coincide with the corresponding terms in Eq.~(12) in the Letter. 

In what follows, we will explicitly check that the Sudakov contributions aforementioned are indeed consistent with the CSS evolution of the quark TMD. To that aim, we follow the standard CSS procedure: in the TMD factorisation, the quark distribution can be defined through subtraction,
\begin{equation}
\mathcal F_{q}^{\rm \,(sub.)}(x,\bt,\mu_F,\zeta_c)=\mathcal F_q^{\rm\, (unsub.)}(x,\bt)\sqrt{\frac{S^{\bar
n,v}(\bt)}{S^{n,\bar n}(\bt)S^{n,v}(\bt)}} \ . \label{jcc}
\end{equation}
The un-subtracted TMD quark distribution is defined as (in momentum space)
 \begin{eqnarray}
\mathcal F_q^{\rm\, (unsub.)}(x,\Pt)&=&\frac{1}{2}\int
        \frac{\der\xi^+\der^2\boldsymbol{\xi}_\perp}{(2\pi)^3}e^{-ix\xi^+P_N^-+i\boldsymbol{\xi}_\perp\cdot\Pt}  \left\langle
P\left|\overline\psi(\xi){\cal L}_{n}^\dagger(\xi)\gamma^-{\cal L}_{n}(0)
        \psi(0)\right|P\right\rangle\ ,\label{tmdun}
\end{eqnarray}
with the gauge link  $ {\cal L}_{n}(\xi) \equiv \exp\left(-ig\int^{-\infty}_0 d\lambda
\, v\cdot A(\lambda n +\xi)\right)$, where the proton is moving along $-\hat z$ direction with large $P_N^-$. \eqn{jcc} also involves the soft factor $S^{v_1,v_2}$ defined as
\begin{equation}
S^{v_1,v_2}(\bt)={\langle 0|{\cal L}_{v_2}^\dagger(\bt) {\cal
L}_{v_1}^\dagger(\bt){\cal L}_{v_1}(0){\cal
L}_{v_2}(0)  |0\rangle   }\, . \label{softg}
\end{equation}
Here,  $\bt$ is the Fourier conjugate variable with respect to the transverse momentum
$\Pt$ and $ \mu_F$ is the factorization scale.  And $\zeta_c^2$ is defined as
$\zeta_c^2=x^2(2v\cdot P)^2/v^2=2(xP_N^-)^2e^{2y_n}$ with $y_n$ being the rapidity cut-off in the
Collins-11 scheme. The second factor represents the soft factor subtraction with $n$ and $\bar n$
as the light-front vectors $n=(0^-,1^+,0_\perp)$, $\bar n=(1^-,0^+,0_\perp)$, whereas $v$ is an
off-light-front $v=(v^-,v^+,0_\perp)$ with $v^+\gg v^-$.

A standard one-loop calculation using  the Collins-11 scheme yields~\cite{Collins:2011zzd,Sun:2013hua} 
\begin{align}\label{eq:F1L}
x\tilde{\mathcal{F}}_q^{\rm \,(sub.)}(x,\bt, \mu_F)&=\frac{\alpha_sC_F}{2\pi}\int_{x}^1\der\xi\,\frac{x}{\xi}f_q\left(\frac{x}{\xi}\right)\left\{\left(-\frac{1}{\epsilon}+\ln\frac{\mu_b^2}{\mu^2}\right)
\frac{1}{C_F}P_{qq}^{(+)}(\xi)+(1-\xi)\right.\nonumber\\
&\left.+\delta(1-\xi)\left[
-\frac{1}{2}\left(\ln\frac{\zeta_c^2}{\mu_b^2}\right)^2
+\frac{3}{2}\ln\frac{\mu_F^2}{\mu_b^2}
+\frac{1}{2}\left(\ln\frac{\zeta_c^2}{\mu^2_F}\right)^2
\right]\right\},
\end{align}
where $\mu_F$ is the factorisation scale and $\zeta_c^2$ has been defined above.

In order to deal with the Sudakov single log proportional to $\alpha_s C_F/(2\pi)\ln(R)$, we finally introduce the soft factor associated with the final state jet,  defined according to
\begin{eqnarray}
\tilde{S}_J(\bt,\mu_F)=\frac{\tilde{S}^R_{n,\bar n}(\bt)}{\sqrt{\tilde{S}_{n,\bar n}(\bt)}}\ ,\label{softfactor}
\end{eqnarray}
where $R$ represents the soft gluon radiation out of the jet cone contribution. 
A one-loop calculation of the above soft factor associated with the jet (similar to that presented in Ref.~\cite{Liu:2018trl}) leads to the following result
\begin{eqnarray}
S_J^{(1)}(P_\perp)=\frac{\alpha_s C_F}{2\pi^2}\frac{1}{P_\perp^2}\ln\frac{1}{R^2}\ ,
\end{eqnarray}
in the transverse momentum space. In $b_\perp$-space, we obtain, to first order,
\begin{eqnarray}
\tilde{S}_J^{(1)}(\bt,\mu_F)=\frac{\alpha_s}{2\pi}\left[-\ln\frac{1}{ R^2}\ln\frac{\mu_F^2}{\mu_b^2}\right]\ .
\end{eqnarray}
From this result, we derive the anomalous dimension at one-loop order,
\begin{eqnarray}
\gamma_s^{(1)}=-C_F\frac{\alpha_s}{2\pi}\ln\frac{1}{ R^2} \ .
\end{eqnarray}
With the above results, we can verify the TMD factorisation at this order,
\begin{eqnarray}
xF_{UU}(x,\bt,\mu_F,\zeta_c)&=&
x\tilde{\mathcal{F}}_{q}^{\textrm{(sub.)}}(x,\bt,\mu_F,\zeta_c) \tilde{S}_J(\bt,\mu_F)H_{\rm TMD}(Q,\mu_F) \ ,\label{facb}
\end{eqnarray}
where to the NLO accuracy of interest, $\tilde{S}_J\simeq 1+\tilde{S}_J^{(1)}$ and the hard factor takes the following form:
\begin{eqnarray}
H_{\rm TMD}^{(1)}&=&\frac{\alpha_sC_F}{2\pi}\left[\frac{1}{2}\ln^2\frac{1}{R^2}+\frac{3}{2}\ln\frac{1}{R^2}-\frac{3}{2}-\frac{2\pi^2}{3}
\right] \ , \label{eq:HT}
\end{eqnarray}
where we have chosen $\mu_F^2=\zeta_c^2=Q^2$ to simplify the final
expression. With these particular choices, which are physically meaningful, it is easy to check that the expression for $xF_{UU}^{(1)}(x,\bt, Q^2)$ predicted by TMD factorisation, \eqn{facb}, at one-loop order coincides with our NLO result for the jet cross-section, as presented in \eqn{euu}. For instance, the last double logarithm within the square brackets in \eqn{eq:F1L} vanishes, while the first two logarithms there account for the double Sudakov logarithm and the Sudakov single log proportional to $3\alpha_s C_F/(4\pi)$ in Eq.\,\eqref{euu}.
One can similarly check all the other terms in the r.h.s. of \eqn{euu}. Importantly,
the hard factor \eqref{eq:HT} does not contain any large Sudakov logarithm of the form $\ln(Q^2/\mu_b^2)$, which means that the CSS evolution of the quark TMD distribution correctly accounts for the Sudakov logarithms which were initially present in the NLO correction given by Eq.\,\eqref{euu}. Since the value of the coefficients of these Sudakov logarithms in Eq.\,\eqref{euu} is a consequence of the choice of jet definition, we obtain again, but this time in the moderate $x$ regime, that our jet definition is consistent with TMD factorisation and CSS+DGLAP evolution.

To conclude this discussion, let us recall that the NLO corrections to jet production in the valence quark picture are unable to distinguish between different values of the parameter $\beta$ in the $B$ class of clustering conditions,  \eqn{eq:cluster-DIS}. The $\beta$--dependence should however become visible at NNLO, or two loop, order. Our previous calculation using sea quarks in the dipole picture predicts that the only acceptable value (in the sense of the conformity with TMD factorisation) is $\beta=0$. 

\end{widetext}

 \end{document}